\documentclass[11pt]{article}
\usepackage[a4paper, total={6.5in, 8in}]{geometry}

\usepackage{amsmath, amssymb, amsthm}
\usepackage{hyperref}
\usepackage{braket}
\usepackage{url}

\usepackage{color}
\usepackage{xcolor}

\newtheorem{theorem}{Theorem}
\newtheorem{lemma}{Lemma}
\newtheorem{proposition}{Proposition}
\newtheorem{corollary}{Corollary}

\theoremstyle{definition}

\theoremstyle{remark}

\DeclareMathOperator{\arctanh}{\mathrm{arctanh}}

\title{Convergence Analysis of the Quantum Blahut-Arimoto Algorithm and the Contraction Coefficient of Qubit Amplitude-Damping Channels}

\author{
Junke Han \quad Yonglong Li \\
School of Mathematics and Statistics, Xi'an Jiaotong University \\
\texttt{hanjunke@stu.xjtu.edu.cn\quad liyonglong@xjtu.edu.cn}
}

\date{\today}

\providecommand{\Pxz}{\mathbb{P}_{xz}}
\providecommand{\Py}{\mathbb{P}_{y}}

\begin{document}

\maketitle

\begin{abstract}
    In this work, we investigate the convergence rate of the quantum Blahut-Arimoto (BA) algorithm proposed for computing the coherent information of less noisy quantum channels. Previous work established exponential convergence under certain restrictive assumptions on the channel. Using a Taylor expansion around a positive definite optimizer, we show that, for an injective degradable channel whose degrading channel has a relative-entropy contraction coefficient strictly less than one, the algorithm converges at an exponential rate after sufficiently many iterations. We further verify this condition for qubit amplitude-damping (AD) channels in the relevant parameter regime. 
    
    In addition, by analyzing the Bogoliubov-Kubo-Mori (BKM) metric on the Bloch ball, we derive an explicit characterization of the relative-entropy contraction coefficient of qubit AD channels. The proposed approach is further extended to calculate the relative contraction coefficient and the relative expansion coefficient of two qubit AD channels.
\end{abstract}


\tableofcontents

\section{Introduction}
Since Shannon's foundational work~\cite{ref1}, channel capacity has been one of the central quantities in information theory. For a discrete memoryless classical channel $P_{Y|X}$, the capacity is given by
\begin{align}
    C(P_{Y|X})=\max_{p_X} I(X;Y),
\end{align}
where the maximization is over input distributions. Although this formula gives a complete variational characterization, it rarely leads to a closed-form expression. Thus, the computation of channel capacity has been a basic and persistent numerical problem in information theory.

One of the most important methods for this problem is the Blahut-Arimoto algorithm, independently introduced by Blahut and Arimoto for the computation of channel capacities and rate-distortion functions \cite{ref2,ref3}.
The algorithm rewrites the capacity problem as an alternating optimization problem over an input distribution and an auxiliary posterior distribution. This structure leads to a simple monotone iteration which is often much faster in practice than generic first-order methods. Existing analyses have established convergence of the quantum BA algorithm and provided sublinear worst-case bounds on the gap between the optimal value and the value obtained at the $t$-th iteration, while Arimoto's original work also revealed a sharper phenomenon: under suitable assumptions, the iteration converges exponentially fast once it is sufficiently close to the capacity-achieving distribution.
This local nature of the exponential convergence result is important. It suggests that the asymptotic behavior of Blahut-Arimoto type algorithms is controlled not only by global variational inequalities, but also by the local geometry of the objective near the optimizer.
The convergence speed of the classical iteration and its accelerations have been analyzed further in~\cite{ref4,ref5}.

The Blahut-Arimoto algorithm has since been reinterpreted from several complementary viewpoints. Csisz\'ar and Tusn\'ady's alternating minimization framework connects it to information projections and information geometry \cite{ref6}.
Later work also interpreted accelerated variants through natural-gradient and proximal-point methods, where relative entropy plays the role of the intrinsic divergence governing the iteration~\cite{ref7,ref8}.
More recently, Bregman proximal and mirror-descent perspectives have further clarified why the algorithm is well adapted to entropic optimization problems \cite{ref9}.
These interpretations are particularly relevant for quantum generalizations, where the ordinary Euclidean geometry of matrices is poorly suited to objectives involving von Neumann entropy and matrix logarithms.

Quantum information theory introduces several different notions of channel capacity, depending on the type of information being transmitted and on the auxiliary resources available, such as pre-shared entanglement~\cite{ref10}.
For classical communication over classical-quantum channels, the Holevo-Schumacher-Westmoreland theorem identifies the operational rate with an optimization of the Holevo quantity~\cite{ref11,ref12}.
For quantum communication, the fundamental quantity is the coherent information
\begin{align}
    I_{\mathrm{coh}}(\rho,\Phi)
    =
    S\left(\Phi(\rho)\right)-S\left(\Phi^{c}(\rho)\right),
\end{align}
where $\Phi$ is a quantum channel and $\Phi^{c}$ is a complementary channel of $\Phi$.

The Lloyd-Shor-Devetak theorem expresses the quantum capacity as the regularization of the optimized coherent information~\cite{ref13,ref14,ref15}.
In general this regularization is unavoidable, reflecting the possible non-additivity of coherent information. However, for important classes of channels, including degradable channels and more generally certain less-noisy channels, coherent information is additive, and the capacity problem reduces to the finite-dimensional optimization of $I_{\mathrm{coh}}(\rho,\Phi)$ \cite{ref16,ref17,ref18}.
This makes the efficient computation of coherent information a natural quantum analogue of the classical capacity-computation problem.
Besides Blahut-Arimoto type iterations, other numerical approaches to entropic optimization problems of this kind include convex-optimization methods, approximation schemes based on smoothing and duality, and semi-definite approximations of the matrix logarithm~\cite{ref19,ref20,ref21,ref22}.

Blahut-Arimoto type methods have also been developed in the quantum setting. Early work of Nagaoka initiated the use of Arimoto-Blahut ideas for quantum channel-capacity computations~\cite{ref23,ref24}.
More recently, Ramakrishnan, Iten, Scholz, and Berta developed a unified family of quantum Blahut-Arimoto algorithms for several entropic channel quantities, including the Holevo quantity, the coherent information of less noisy channels, the thermodynamic capacity, and the quantum mutual information~\cite{ref25}. Related algorithms for classical-quantum channels and for quantum information-bottleneck problems were developed in~\cite{ref26,ref27}.
For coherent information, the iteration of~\cite{ref25}, with acceleration parameter $\gamma=1$, has the form of a matrix-exponential update,
\begin{align}
    \rho^{(t+1)}
    =
    \frac{1}{Z(t+1)}
    \exp\!\left(
        \log \rho^{(t)} + \mathcal{F}_{\Phi}(\rho^{(t)})
    \right).
\end{align}
This update is the natural noncommutative analogue of the classical Blahut-Arimoto iteration: probability vectors are replaced by density operators, relative entropy is replaced by quantum relative entropy, and ordinary logarithms are replaced by matrix logarithms.

The convergence guarantees for quantum Blahut-Arimoto algorithms obtained in~\cite{ref25} show that the algorithm converges exponentially under a uniform lower-bound condition on relative entropy. In particular, this requires the relative-entropy expansion coefficient of the channel to be strictly positive. This requirement can be restrictive, since many non-unitary quantum channels have zero expansion coefficient, including channels whose input dimension is larger than their output dimension and whose Kraus rank exceeds one~\cite{ref28}. Consequently, a global expansion-coefficient condition cannot explain the fast convergence observed in many examples. This situation is analogous to the classical Blahut-Arimoto algorithm: the practical exponential behavior is not necessarily a consequence of a global contraction principle, but may instead arise from the local structure of the iteration around the optimizer. This motivates us to study the asymptotic behavior of the algorithm near the optimizer rather than imposing a uniform condition over the entire state space. 

The first goal of this paper is to establish a local exponential convergence result for the quantum Blahut-Arimoto algorithm for coherent information. We first prove that the iterates converge to the unique optimizer. Assuming that this optimizer is positive definite, a Taylor expansion of the quantum $\chi^{2}$-divergence gives a local quadratic upper bound on the relative entropy to the optimizer. Combining this bound with injectivity and a strong data-processing inequality for the degrading channel yields a contraction estimate along the iteration and hence exponential convergence after finitely many iterates.

For degradable channels, this local analysis naturally leads to contraction coefficients. If $\Phi$ is degradable, then its complementary channel can be written as
\begin{align}
    \Phi^{c} = \mathcal{D} \circ \Phi
\end{align}
for a degrading channel $\mathcal{D}$.
The coherent information can then be viewed as the difference between the entropy of the receiver's output and the entropy after the degrading map. In the second-order expansion of this difference, the relevant comparison is precisely how much the degrading map contracts relative entropy, or equivalently its infinitesimal quadratic form. Thus, the contraction coefficient of the degrading map is not an independent object inserted into the proof. Instead it appears intrinsically as the local curvature term controlling the stability of the Blahut-Arimoto update. This observation motivates the second part of the paper.

The contraction coefficient of a quantum channel $\Phi$ with respect to quantum relative entropy is defined by
\begin{align}
    \eta(\Phi)
    =
    \sup_{\rho\neq \sigma,\rho\ll\sigma}
    \frac{D(\Phi(\rho)\Vert\Phi(\sigma))}
         {D(\rho\Vert\sigma)}.
\end{align}
It refines the data-processing inequality by identifying the best constant in the inequality
\begin{align}
    D(\Phi(\rho)\Vert\Phi(\sigma))
    \leq
    \eta(\Phi) D(\rho\Vert\sigma).
\end{align}
When $\eta(\Phi)<1$, the channel satisfies a strong data-processing inequality, meaning that it uniformly decreases distinguishability. Such coefficients have a long history in classical information theory and have become increasingly important in quantum information theory, where they are connected to strong data-processing inequalities, channel comparison, hypercontractivity, mixing, and partial orders on quantum channels~\cite{ref29,ref30,ref31,ref32,ref33,ref34}.

A powerful way to study relative-entropy contraction is through the BKM metric~\cite{ref32}. Quantum relative entropy has a second-order expansion
\begin{align}
    D(\rho+\varepsilon X\Vert \rho)
    =
    \frac{\varepsilon^2}{2}
    \langle X,\Omega_\rho(X)\rangle
    +o(\varepsilon^2),
\end{align}
where $\rho$ is positive definite, $X$ is traceless and Hermitian, and $\Omega_\rho$ is the BKM operator.
Thus, infinitesimal relative-entropy contraction can be studied by comparing BKM quadratic forms before and after a channel. For quantum channels, this local Riemannian viewpoint is especially useful because the global optimization over two density matrices is generally non-convex and analytically difficult. In the qubit case, the Bloch-ball representation makes the BKM metric explicit and allows symmetries of the channel to reduce the problem to a much lower-dimensional optimization.

We apply this method to the qubit AD channel. This channel is one of the basic non-unital noise models in quantum information theory and describes relaxation from an excited state to a ground state. It is also a natural test case for quantum Blahut-Arimoto algorithms: it is simple enough to permit explicit analysis, but it remains genuinely noncommutative and lacks the full symmetry like depolarizing channels.

For damping parameter $p\in[0,1]$, the channel acts as
\begin{align}
    \mathcal{A}_{p}
    \begin{pmatrix}
        \rho_{00} & \rho_{01}\\
        \rho_{10} & \rho_{11}
    \end{pmatrix}
    =
    \begin{pmatrix}
        \rho_{00}+p\rho_{11}
        &
        \sqrt{1-p}\,\rho_{01}\\
        \sqrt{1-p}\,\rho_{10}
        &
        (1-p)\rho_{11}
    \end{pmatrix}.
\end{align}
For $p\leq 1/2$ it is degradable, and its degrading map is again an AD channel with an explicitly related parameter~\cite{ref35,ref36}. Understanding the contraction coefficient of AD channels gives concrete sufficient conditions for the exponential convergence of the quantum Blahut-Arimoto algorithm in this important example.

The second main contribution of this paper is an explicit characterization of relative-entropy contraction for AD channels. Using the axial symmetry of the channel on the Bloch ball and the explicit form of the BKM metric for qubits, we reduce the optimization defining the contraction coefficient to a single-variable problem. We then extend the same method to compare two AD channels. More precisely, for two channels $\mathcal{A}_{p_1}$ and $\mathcal{A}_{p_2}$, we study the relative contraction coefficient
\begin{align}
    \eta(\mathcal{A}_{p_1},\mathcal{A}_{p_2})
    =
    \sup_{\rho\neq\sigma,\rho\ll\sigma}
    \frac{
        D(\mathcal{A}_{p_1}(\rho)\Vert
          \mathcal{A}_{p_1}(\sigma))
    }{
        D(\mathcal{A}_{p_2}(\rho)\Vert
          \mathcal{A}_{p_2}(\sigma))
    },
\end{align}
as well as the corresponding relative expansion coefficient obtained by replacing the supremum with the infimum.
These quantities measure how one AD channel contracts or preserves distinguishability relative to another. They are naturally connected to reverse data-processing inequalities and to less-noisy comparisons between quantum channels.

Our results answer open questions raised in recent work of Belzig, Gao, Smith, and Wu on reverse-type data-processing inequalities~\cite{ref28}. In that work, relative contraction and expansion coefficients were introduced as a way to obtain meaningful reverse inequalities even when the ordinary expansion coefficient is zero. We show that the remaining symmetry of the AD channel is nevertheless sufficient to solve the optimization problems exactly after passing to the BKM-metric formulation.

The two parts of the paper are tied together by a common principle: the local geometry of quantum relative entropy controls both the convergence of quantum Blahut-Arimoto algorithms and the contraction behavior of quantum channels. In the convergence problem, a Taylor expansion of the quantum $\chi^{2}$-divergence near a positive definite optimizer provides the local quadratic bound needed to establish exponential convergence.
In the contraction problem, the same second-order expansion of relative entropy produces the BKM metric, whose comparison under AD channels determines optimal contraction and relative contraction constants. Thus, although algorithmic convergence and reverse data processing appear at first to be separate topics, both are governed by the same local analytic structure.

The paper is organized as follows. In Section~\ref{sec:prelim}, we recall the necessary background on quantum channels, coherent information, degradability, quantum relative entropy, the BKM metric, and the quantum Blahut-Arimoto algorithm.
In Section~\ref{sec:main}, we state our main results. 
In Section~\ref{sec:proof}, we prove the convergence results for the quantum Blahut-Arimoto algorithm and verify their assumptions for the qubit dephasing and AD channels.
In Section~\ref{sec:contractionproof}, we prove the results on the contraction, relative contraction and relative expansion coefficients of AD channels.
Technical computations are collected in the appendices.

\section{Preliminaries}\label{sec:prelim}

In this section, we fix notations and recall several basic notions that will be used throughout the paper. 

\subsection{Quantum channels and the coherent information}

Throughout this paper, all Hilbert spaces are assumed to be finite-dimensional. Unless otherwise specified, all logarithms are natural.
We denote Hilbert spaces by $\mathcal{X},\mathcal{Y},\mathcal{Z}$, and so on. 
The space of linear operators from $\mathcal{X}$ to $\mathcal{Y}$ is denoted by $\mathcal{L}(\mathcal{X},\mathcal{Y})$, and we write $\mathcal{L}(\mathcal{X})$ as a shorthand for $\mathcal{L}(\mathcal{X},\mathcal{X})$.
The adjoint (conjugate transpose) of an operator $A$ is denoted by $A^{\dagger}$.
We equip $\mathcal{L}(\mathcal{X})$ with the Hilbert-Schmidt inner product $\langle A,B \rangle=\mathrm{Tr}(A^{\dagger}B)$, and we write $\|A\|_{2}=\langle A,A\rangle^{1/2}$, $\|A\|_{1}=\mathrm{Tr}|A|$ and $\|A\|_{\infty}$ for the Hilbert-Schmidt norm, the trace norm and the operator norm, respectively.
The identity operator on $\mathcal{X}$ is denoted by $I_{\mathcal{X}}$, or simply by $I$.

The set of density operators on $\mathcal{X}$ is $\mathcal{D}(\mathcal{X})=\{ \rho \in \mathcal{L}(\mathcal{X}): \rho \geq 0,\ \mathrm{Tr}(\rho)=1 \}$.
The von Neumann entropy of a density operator $\rho \in \mathcal{D}(\mathcal{X})$ is defined as
\begin{align}
    S(\rho) = -\mathrm{Tr}(\rho \log \rho).
\end{align}
We write $\mathcal{T}(\mathcal{X},\mathcal{Y})$ for the space of linear maps from $\mathcal{L}(\mathcal{X})$ to $\mathcal{L}(\mathcal{Y})$, and write $\mathcal{T}(\mathcal{X})=\mathcal{T}(\mathcal{X},\mathcal{X})$.
A linear map $\Phi \in \mathcal{T}(\mathcal{X},\mathcal{Y})$ is called a \emph{quantum channel} if it is completely positive and trace-preserving.
The adjoint $\Phi^{\dagger}\in\mathcal{T}(\mathcal{Y},\mathcal{X})$ of $\Phi$ is defined by $\langle \Phi^{\dagger}(B),A\rangle=\langle B,\Phi(A)\rangle$ for all $A\in\mathcal{L}(\mathcal{X})$ and $B\in\mathcal{L}(\mathcal{Y})$.

For two density operators $\rho$ and $\sigma$, the \emph{quantum relative entropy} (see, e.g.,~\cite{ref37} for its operational meaning) is defined by
\begin{align}
    D(\rho \| \sigma)
    =
    \begin{cases}
    \mathrm{Tr}\left[\rho(\log\rho-\log\sigma)\right],
    & \text{if } \mathrm{supp}(\rho) \subseteq \mathrm{supp}(\sigma), \\
    +\infty,
    & \text{otherwise}.
    \end{cases}
\end{align}
Equivalently, the support condition may be written as $\mathrm{ker}(\sigma) \subseteq \mathrm{ker}(\rho)$, and we denote it by $\rho \ll \sigma$.
For two density operators $\rho$ and $\sigma$, the \emph{quantum $\chi^{2}$-divergence} is defined by
\begin{align}
    \chi^{2} (\rho \| \sigma)
    = \begin{cases}
    \mathrm{Tr}\left(\rho^{2} \sigma^{-1}\right)-1 ,
    & \text{if } \mathrm{supp}(\rho) \subseteq \mathrm{supp}(\sigma), \\
    +\infty,
    & \text{otherwise},
    \end{cases}
\end{align}
where $\sigma^{-1}$ is understood as the inverse on $\mathrm{supp}(\sigma)$. From~\cite{ref38}, we have the following inequality:
\begin{align}
    D(\rho \| \sigma)
    \leq
    \log \left( 1+\chi^{2}(\rho \| \sigma) \right)
    \leq
    \chi^{2}(\rho \| \sigma) .\label{eq:chi}
\end{align}
%The first inequality follows from the monotonicity in $\alpha$ of the Petz R\'enyi divergence $\bar{D}_{\alpha}$, defined by $\bar{D}_{\alpha}(\rho\|\sigma)=\frac{1}{\alpha-1}\log\mathrm{Tr}(\rho^{\alpha}\sigma^{1-\alpha})$, since $\bar{D}_{1}=D$ and $\bar{D}_{2}=\log(1+\chi^{2})$.}

Let $\Phi \in \mathcal{T}(\mathcal{X},\mathcal{Y})$ be a quantum channel. By the Stinespring dilation theorem, there exist a finite-dimensional environment system $\mathcal{E}$ and an isometry $V: \mathcal{X} \rightarrow \mathcal{Y} \otimes \mathcal{E}$ such that
\begin{align}
    \Phi(\rho)
    =
    \mathrm{Tr}_{\mathcal{E}}
    \left( V \rho V^{\dagger} \right),
    \quad
    \rho \in \mathcal{D}(\mathcal{X}) ,
\end{align}
where $\mathrm{Tr}_{\mathcal{E}}$ denotes the partial trace over the environment system $\mathcal{E}$. The \emph{complementary channel} of $\Phi$ associated with this Stinespring representation is defined by
\begin{align}
    \Phi^{c}(\rho)
    =
    \mathrm{Tr}_{\mathcal{Y}}
    \left( V \rho V^{\dagger} \right),
    \quad
    \rho \in \mathcal{D}(\mathcal{X}) .
\end{align}

A quantum channel $\Phi$ is said to be \emph{less noisy} if the private capacity of its complementary channel $\Phi^{c}$ is zero~\cite{ref17}.
The channel $\Phi$ is called \emph{degradable} if there exists a quantum channel $\mathcal{D}\in\mathcal{T}(\mathcal{Y},\mathcal{E})$ such that $\mathcal{D} \circ \Phi = \Phi^{c}$~\cite{ref16,ref39}.
In this case, $\mathcal{D}$ is called a degrading channel of $\Phi$.
It is well known that every degradable channel is less noisy~\cite{ref17}.

The \emph{coherent information} of $\Phi$ is defined by
\begin{align}
    I_{\mathrm{coh}}(\Phi)
    :=
    \max_{\rho \in \mathcal{D}(\mathcal{X})}
    I_{\mathrm{coh}}(\rho,\Phi),
\end{align}
where
\begin{align}
    I_{\mathrm{coh}}(\rho,\Phi)
    =S\left(\Phi(\rho)\right)-S\left(\Phi^{c}(\rho)\right)
\end{align}
denotes the coherent information of $\Phi$ with respect to the input state $\rho$.

\subsection{BKM metric and relative entropy coefficients}
\label{sec:BKMsub}

In this subsection, we recall the BKM metric and its connection with relative entropy coefficients.
This will be the main tool for computing contraction coefficients and relative contraction coefficients later.

The \emph{BKM metric} $g_{\sigma}(X)$ of an operator $X \in \mathcal{L}(\mathcal{X})$ at a density operator $\sigma \in \mathcal{D}(\mathcal{X})$ is defined as~\cite{ref28,ref32}
\begin{align}
    g_{\sigma}(X)
    =
    \begin{cases}
    \mathrm{Tr} \left( \int_{0}^{\infty} X^{\dagger}(\sigma + rI)^{-1}X(\sigma + rI)^{-1}dr \right),
    & \text{if } \mathrm{supp}(X) \subseteq \mathrm{supp}(\sigma), \\
    +\infty,
    & \text{otherwise}.
    \end{cases}
\end{align}
Let $\Phi$ be a quantum channel.

The \emph{contraction coefficient} $ \eta(\Phi)$ and the \emph{expansion coefficient} $\check{\eta}(\Phi)$ of $\Phi$ with respect to quantum relative entropy are defined respectively by
\begin{align}
    \eta(\Phi)
    &:=
    \sup_{\substack{\rho \ll \sigma \\ \rho \neq \sigma}}
    \frac{D \left( \Phi(\rho) \| \Phi(\sigma) \right)}
    {D(\rho \| \sigma)} 
    \quad\mbox{and}\quad
    \check{\eta}(\Phi)
    :=
    \inf_{\substack{\rho \ll \sigma \\ \rho \neq \sigma}}
    \frac{D \left( \Phi(\rho) \| \Phi(\sigma) \right)}
    {D(\rho \| \sigma)} .
\end{align}
Since $0<D(\rho \| \sigma)<\infty$ whenever $\rho\ll\sigma$ and $\rho \neq \sigma$, the above ratios are well defined.
By the data processing inequality, one always has
\begin{align}
    0
    \leq
    \check{\eta}(\Phi)
    \leq
    \eta(\Phi)
    \leq
    1 .
\end{align}

The corresponding coefficients with respect to the BKM metric are defined by taking the infinitesimal form of the above ratios. More precisely, the \emph{BKM metric contraction coefficient} and the \emph{BKM metric expansion coefficient} of $\Phi$ are defined respectively by
\begin{align}
    \eta_{\mathrm{BKM}}(\Phi)
    &:=
    \sup_{\substack{\sigma\in\mathcal{D}(\mathcal{X})\\
        X\in\mathcal{T}_{\sigma},\ X\neq0}}
    \frac{g_{\Phi(\sigma)}\bigl(\Phi(X)\bigr)}{g_{\sigma}(X)},
\end{align}
and
\begin{align}
    \check{\eta}_{\mathrm{BKM}}(\Phi)
    &:=
    \inf_{\substack{\sigma\in\mathcal{D}(\mathcal{X})\\ X\in\mathcal{T}_{\sigma},\ X\neq0}}
    \frac{g_{\Phi(\sigma)}\bigl(\Phi(X)\bigr)}{g_{\sigma}(X)},
\end{align}
where $\mathcal{T}_{\sigma}:=\{X\in\mathcal{L}(\mathcal{X}): X=X^{\dagger},\mathrm{Tr}(X)=0,\mathrm{supp}(X)\subseteq\mathrm{supp}(\sigma)\}$.

Then we introduce relative versions of contraction and expansion coefficients. Let $\Phi$ and $\Psi$
be two quantum channels with the same input system. The \emph{relative contraction coefficient} and the \emph{relative expansion coefficient} of $\Phi$ with respect to $\Psi$ are defined respectively by
\begin{align}
    \eta(\Phi,\Psi)
    :=
    \sup_{\substack{\rho \ll \sigma \\ \rho \neq \sigma}}
    \frac{D \left( \Phi(\rho) \| \Phi(\sigma) \right)}
    {D \left( \Psi(\rho) \| \Psi(\sigma) \right)} 
\end{align}
and
\begin{align}
    \check{\eta}(\Phi,\Psi)
    :=
    \inf_{\substack{\rho \ll \sigma \\ \rho \neq \sigma}}
    \frac{D \left( \Phi(\rho) \| \Phi(\sigma) \right)}
    {D \left( \Psi(\rho) \| \Psi(\sigma) \right)}.
\end{align}
Whenever both quantities are well defined, we have the reciprocal relation
\begin{align}
    \check{\eta}(\Phi,\Psi)
    =
    \frac{1}{\eta(\Psi,\Phi)} .
\end{align}

Similarly, the \emph{BKM metric relative contraction coefficient} of $\Phi$ with respect to $\Psi$ is defined by
\begin{align}
    \eta_{\mathrm{BKM}}(\Phi,\Psi)
    &:=
    \sup_{\substack{
        \sigma\in\mathcal{D}(\mathcal{X})\\
        X\in\mathcal{T}_{\sigma},\ X\neq0}}
    \frac{g_{\Phi(\sigma)}\bigl(\Phi(X)\bigr)}{g_{\Psi(\sigma)}\bigl(\Psi(X)\bigr) },
\end{align}
while the \emph{BKM metric relative expansion coefficient} of $\Phi$ with respect to $\Psi$ is defined by
\begin{align}
    \check{\eta}_{\mathrm{BKM}}(\Phi,\Psi)
    &:=
    \inf_{\substack{
        \sigma\in\mathcal{D}(\mathcal{X})\\
        X\in\mathcal{T}_{\sigma},\ X\neq0}}
    \frac{g_{\Phi(\sigma)}\bigl(\Phi(X)\bigr)}{g_{\Psi(\sigma)}\bigl(\Psi(X)\bigr)}.
\end{align}
Whenever both quantities are well defined, we also have
\begin{align}
    \check{\eta}_{\mathrm{BKM}}(\Phi,\Psi)
    =
    \frac{1}{\eta_{\mathrm{BKM}}(\Psi,\Phi)} .
\end{align}

The following lemma, a one-sided version of~\cite[Lemma~4.1]{ref28}, is the main tool for studying these coefficients.

\begin{lemma}[Belzig \emph{et al.}, Lemma \uppercase\expandafter{\romannumeral4}.1]
\label{lemma:BKM}

Let $\Phi\in\mathcal{T}(\mathcal{X},\mathcal{Y})$ and $\Psi\in\mathcal{T}(\mathcal{X},\mathcal{Z})$ be quantum channels and let $c\geq0$.
The following two statements are equivalent:

(1) For all density operators $\rho$ and $\sigma$ with $\rho\ll\sigma$,
\begin{align}
    D \left( \Psi(\rho) \| \Psi(\sigma) \right)
    \leq
    c\, D \left( \Phi(\rho) \| \Phi(\sigma) \right) .
\end{align}

(2) For all density operators $\sigma$ and all traceless Hermitian operators $X$ with $\mathrm{supp}(X)\subseteq\mathrm{supp}(\sigma)$,
\begin{align}
    g_{\Psi(\sigma)} \left( \Psi(X) \right)
    \leq
    c\, g_{\Phi(\sigma)} \left( \Phi(X) \right) .
\end{align}
\end{lemma}

As a consequence of Lemma~\ref{lemma:BKM} (applied with the roles of $\Phi$ and $\Psi$ interchanged), the relative entropy coefficients coincide with the corresponding BKM metric coefficients:
\begin{align}
    \eta(\Phi,\Psi)
    =
    \eta_{\mathrm{BKM}}(\Phi,\Psi),
    \quad
    \check{\eta}(\Phi,\Psi)
    =
    \check{\eta}_{\mathrm{BKM}}(\Phi,\Psi).
\end{align}
In particular, taking $\Psi$ to be the identity channel gives
\begin{align}
    \eta(\Phi)=\eta_{\mathrm{BKM}}(\Phi),
    \quad
    \check{\eta}(\Phi)=\check{\eta}_{\mathrm{BKM}}(\Phi).
\end{align}

We now specialize to qubit systems. Let $\vec{\sigma}=(\sigma_{x},\sigma_{y},\sigma_{z})$ denote the vector of Pauli matrices.
Every qubit density operator admits a Bloch representation of the form
\begin{align}
    \rho_{\vec{w}} = \frac{1}{2}( I_{2}+\vec{w} \cdot \vec{\sigma}),
    \quad
    |\vec{w}| \leq 1 ,
\end{align}
where $\vec{w}\in\mathbb{R}^{3}$ is the Bloch vector of $\rho_{\vec{w}}$.  We shall also use the same Bloch coordinates for traceless Hermitian perturbations. More precisely, every traceless Hermitian operator can be written uniquely as
\begin{align}
    X_{\vec{v}}=\vec{v} \cdot \vec{\sigma},
    \quad
    \vec{v} \in \mathbb{R}^{3} .
\end{align}
For later computations, it is convenient to combine the above representation with the Bloch representation of qubit channels. If $\Phi$ is a qubit channel, then there exist a real matrix $T_{\Phi} \in M_{3}(\mathbb{R})$ and a vector $\vec{t}_{\Phi} \in \mathbb{R}^{3}$ such that
\begin{align}
    \Phi(\rho_{\vec{w}})
    &=
    \frac{1}{2}
    \left(I_{2}+(T_{\Phi}\vec{w}+\vec{t}_{\Phi}) \cdot \vec{\sigma}\right) 
\end{align}
and
\begin{align}
    \Phi(X_{\vec{v}})
    &=
    (T_{\Phi}\vec{v}) \cdot \vec{\sigma} .
\end{align}

Let $f : [0,1] \to \mathbb{R}$ be defined by
\begin{align}
    f(x) :=
    \frac{1-x^{2}}{2x}\log\frac{1+x}{1-x}
    =\frac{(1-x^{2})\arctanh(x)}{x},
    \quad x \in (0,1),
\end{align}
with $f(0):=1$ and $f(1):=0$ by continuity.
Expanding $\arctanh(x)=\sum\limits_{k\geq0}\frac{x^{2k+1}}{2k+1}$ gives, for $x\in[0,1)$,

\begin{align}
    \frac{\arctanh(x)}{x}=\sum_{k\geq 0}\frac{x^{2k}}{2k+1},
    \qquad
    f(x)=1-\sum_{k\geq 1}\frac{2x^{2k}}{(2k-1)(2k+1)} .
    \label{eq:fseries}
\end{align}
Hence $x\mapsto\arctanh(x)/x$ is increasing on $[0,1)$,
and $f$ is strictly decreasing on $[0,1]$ with $0<f(x)<1$ for $x\in(0,1)$. Equivalently, $(1-x^{2})\arctanh(x)<x$ for $x\in(0,1)$.
Throughout the paper, $\arctanh(x)/x$ and $x/\arctanh(x)$ are assigned the value $1$ at $x=0$ by continuity. These facts are used repeatedly below.

The following explicit formula~\cite{ref28} will be used later.

\begin{lemma}[Belzig \emph{et al.}, Lemma \uppercase\expandafter{\romannumeral4}.4]\label{lemma:formula}

Let $\rho_{\vec{w}}=\frac{1}{2}\left(I_{2}+\vec{w}\cdot\vec{\sigma}\right)$ and $X_{\vec{v}}=\vec{v}\cdot\vec{\sigma}$ with $|\vec{w}|<1$.
Then we have 
\begin{align}
    g_{\vec{w}}(\vec{v})
    &\triangleq g_{\rho_{\vec{w}}}(X_{\vec{v}})  \\
    &= \frac{4|\vec{v}|^{2}}{1-|\vec{w}|^{2}} \left[ \frac{|\vec{w}\cdot \vec{v}|^{2}}{|\vec{w}|^{2}|\vec{v}|^{2}} \left( 1-f(|\vec{w}|) \right) + f(|\vec{w}|) \right].
\end{align}
For $\vec{w}=\vec{0}$ the first term in the bracket is understood by continuity as $0$, and $g_{\vec{0}}(\vec{v})=4|\vec{v}|^{2}$. For $\vec{v}=\vec{0}$ set $g_{\vec{w}}(\vec{0})=0$ directly, without evaluating the displayed quotient.

Let $\vec{w}_{\Phi}=T_{\Phi}\vec{w}+\vec{t}_{\Phi}$ and $\vec{v}_{\Phi}=T_{\Phi}\vec{v}$, and assume $|\vec{w}_{\Phi}|<1$. The same zero-vector conventions apply to the following output formula.
Then 
\begin{align}
    g_{\vec{w}_{\Phi}}(\vec{v}_{\Phi})
    &\triangleq g_{\Phi(\rho_{\vec{w}})}\left(\Phi(X_{\vec{v}})\right)  \\
    &= \frac{4|\vec{v}_{\Phi}|^{2}}{1-|\vec{w}_{\Phi}|^{2}} \left[ \frac{|\vec{w}_{\Phi}\cdot \vec{v}_{\Phi}|^{2}}{|\vec{w}_{\Phi}|^{2}|\vec{v}_{\Phi}|^{2}} \left( 1-f(|\vec{w}_{\Phi}|) \right) + f(|\vec{w}_{\Phi}|) \right] .
\end{align}
\end{lemma}

\subsection{Structure of the quantum Blahut-Arimoto algorithm}

In this subsection, we recall the quantum Blahut-Arimoto algorithm introduced in~\cite{ref25}, which is used to compute the coherent information of less noisy channels.

Let $\Phi$ be a less noisy channel. Define
\begin{align}
    \mathcal{F}_{\Phi}(\rho)
    :=
    (\Phi^{c})^{\dagger}\left(\log \Phi^{c}(\rho)\right) - \Phi^{\dagger}\left(\log \Phi(\rho)\right) .
\end{align}

For density operators $\rho,\sigma \in \mathcal{D}(\mathcal{X})$ with $\rho \ll \sigma$, let
\begin{align}
    J_{\gamma}(\rho,\sigma,\Phi)
    :=
    -\gamma D(\rho \| \sigma)
    +
    \mathrm{Tr} \left[ \rho\, \mathcal{F}_{\Phi}(\sigma) \right] ,
\end{align}
where $\gamma>0$ is an acceleration parameter.
A direct computation (cf.~\cite[Section~VI]{ref25}) gives
\begin{align}
    J_{\gamma}(\rho,\sigma,\Phi)
    = I_{\mathrm{coh}}(\rho,\Phi)
    + D(\Phi(\rho)\|\Phi(\sigma)) - D(\Phi^{c}(\rho)\|\Phi^{c}(\sigma))
    - \gamma D(\rho\|\sigma). \label{eq:Jidentity}
\end{align}
Besides, for a less noisy channel $\Phi$, one has~\cite{ref25}
\begin{align}
    D \left( \Phi(\rho) \| \Phi(\sigma) \right)
    -
    D \left( \Phi^{c}(\rho) \| \Phi^{c}(\sigma) \right)
    \geq 0
\end{align}
for all admissible pairs $\rho,\sigma$.

Following~\cite{ref25}, let
\begin{align}
    \zeta_{\mathrm{con}}(\Phi)
    :=
    \sup_{\substack{\rho \ll \sigma \\ \rho \neq \sigma}}
    \frac
    {D \left( \Phi(\rho) \| \Phi(\sigma) \right)
    - D \left( \Phi^{c}(\rho) \| \Phi^{c}(\sigma) \right)}
    {D(\rho\|\sigma)}
    \leq 1
\end{align}
and choose $\gamma$ such that
\begin{align}
    \zeta_{\mathrm{con}}(\Phi)
    \leq
    \gamma
    \leq
    1 .
\end{align}
With this choice, it is easy to verify from~\eqref{eq:Jidentity} that $J_{\gamma}(\rho,\sigma,\Phi) \leq J_{\gamma}(\rho,\rho,\Phi)=I_{\mathrm{coh}}(\rho,\Phi)$ and therefore
\begin{align}
    I_{\mathrm{coh}}(\Phi)
    &=
    \max_{\rho\in \mathcal{D}(\mathcal{X})}\max_{\substack{\sigma \in \mathcal{D}(\mathcal{X}): \rho \ll \sigma}}
    J_{\gamma}(\rho,\sigma,\Phi) \\
    &=
    \max_{\substack{\rho \in \mathcal{D}(\mathcal{X})} }
    J_{\gamma}(\rho,\rho,\Phi) .
\end{align}

The algorithm proposed in~\cite{ref25} consists of the following steps:

\begin{itemize}
\item[(1)] Initially, choose a full-rank density operator $\rho^{(1)} \in \mathcal{D}(\mathcal{X})$.
In practice, the maximally mixed state $I_{\mathcal{X}}/\dim(\mathcal{X})$ is a natural choice.

\item[(2)] For $t \in \mathbb{N}^{+}$, update
\begin{align}
    \rho^{(t+1)}
    =
    \frac{1}{Z(t+1)}
    \exp\left[\log\rho^{(t)}+\frac{1}{\gamma}\mathcal{F}_{\Phi}\left(\rho^{(t)}\right)\right] ,
\end{align}
where the normalization constant is given by
\begin{align}
    Z(t+1)=\mathrm{Tr}
    \left\{\exp\left[\log\rho^{(t)}+\frac{1}{\gamma}\mathcal{F}_{\Phi}(\rho^{(t)})\right]\right\} .
\end{align}

\item[(3)] The value generated by the algorithm at the $t$-th step is
\begin{align}
    I_{\mathrm{coh}}^{(t)}(\Phi)
    :=
    J_{\gamma}(\rho^{(t+1)},\rho^{(t)},\Phi) .
\end{align}
\end{itemize}
Let $\rho^{\star}$ be an optimizer of the coherent information, that is,
\begin{align}
    \rho^{\star} \in \arg\max_{\rho \in \mathcal{D}(\mathcal{X})}
    I_{\mathrm{coh}}(\rho,\Phi) .
\end{align}

Ramakrishnan et al.\ prove the following properties (see~\cite[Theorem~III.3 and its proof]{ref25}), which will be used later.

\begin{itemize}
\item[(1)] The sequence $I_{\mathrm{coh}}^{(t)}(\Phi)$ is monotonically increasing in $t$ and converges to $I_{\mathrm{coh}}(\Phi)$.

\item[(2)] For every $t \in \mathbb{N}^{+}$, one has
\begin{align}
    0 \leq I_{\mathrm{coh}}(\Phi)-I_{\mathrm{coh}}^{(t)}(\Phi) \leq
    \gamma \, D(\rho^{\star} \| \rho^{(t)}) . \label{eq:ac}
\end{align}

\item[(3)] It follows from the update formulas that $\rho^{(t)}>0$ for every $t \in \mathbb{N}^{+}$.
\end{itemize}

The convergence of the quantum Blahut-Arimoto algorithm has also been studied by Ramakrishnan et al. In particular, they prove the following result.

\begin{proposition}[Ramakrishnan \emph{et al.}\label{pro:exponential}, Proposition \uppercase\expandafter{\romannumeral3}.6]

Let $\left\{\rho^{(t)}\right\}_{t \geq 1}$ be the sequence generated by the quantum Blahut-Arimoto iteration for computing the coherent information of a less noisy channel $\Phi$.
Suppose that there exists a constant $a\in(0,\gamma]$ such that
\begin{align}
    \zeta_{\mathrm{exp}}(\Phi):=
    \inf_{\substack{\rho \ll \sigma \\ \rho \neq \sigma}}
    \frac
    {D\left(\Phi(\rho)\|\Phi(\sigma)\right) - D\left(\Phi^{c}(\rho)\|\Phi^{c}(\sigma)\right)}
    {D(\rho\|\sigma)}
    =a.
\end{align}
Then the optimizer $\rho^{\star}$ is unique and
\begin{align}
    I_{\mathrm{coh}}(\Phi)-I_{\mathrm{coh}}^{(t+1)}(\Phi)
    \leq
    \gamma\left(1-\frac{a}{\gamma}\right)^{t}D(\rho^{\star}\|\rho^{(1)}).
\end{align}
\end{proposition}

The quantity $\zeta_{\mathrm{exp}}(\Phi)$ is bounded above by the expansion coefficient $\check{\eta}(\Phi)$ of $\Phi$.
However, recent results on reverse data processing inequalities indicate that such a global positivity condition is extremely restrictive~\cite{ref28}.
In particular, for a quantum channel whose input dimension is not smaller than its output dimension, the expansion coefficient must vanish unless the channel is unitary~\cite[Theorem~3.1]{ref28}.
Consequently, one should not expect the assumption of Proposition~\ref{pro:exponential} to hold for generic quantum channels.

Rather than seeking a global expansion coefficient that is uniformly positive over the entire state space, we investigate whether a local expansion phenomenon emerges along the Blahut-Arimoto trajectory.
The key point is that, although the global infimum may be zero, the proof of Proposition~\ref{pro:exponential} only relies on the one-step estimate
\begin{align}
    D\left(\Phi(\rho^{\star})\|\Phi\left(\rho^{(t)}\right)\right) -
    D\left(\Phi^{c}(\rho^{\star})\|\Phi^{c}(\rho^{(t)})\right)
    \geq
    a \, D(\rho^{\star}\|\rho^{(t)}).\label{eqn:dp}
\end{align}
Whenever such an inequality can be established locally around the optimizer, one still obtains a contraction estimate for $D(\rho^{\star}\|\rho^{(t)})$,
which serves as the starting point of our local convergence analysis.
 More specifically, the relative entropy distance to the optimizer contracts at the next step:
\begin{align}
    D(\rho^{\star} \| \rho^{(t+1)})
    \leq
    \left(1-\frac{a}{\gamma}\right)
    D(\rho^{\star} \| \rho^{(t)}) \label{eq:ad},
\end{align}
whenever \eqref{eqn:dp} holds.

\section{Main Results}\label{sec:main}

In this section we state our main results. The first result, presented in Subsection~\ref{sec:BA}, concerns the convergence behavior of the quantum Blahut-Arimoto algorithm for computing the coherent information.
 The second result, presented in Subsection~\ref{sec:AD}, concerns AD channels: we characterize the contraction coefficient of a single AD channel, as well as the relative contraction and relative expansion coefficients between two AD channels.

\subsection{Exponential convergence of the quantum Blahut-Arimoto algorithm}
\label{sec:BA}

In this subsection we establish sufficient conditions under which the quantum Blahut-Arimoto algorithm converges locally at an exponential rate.   We then verify that these conditions are satisfied by the qubit dephasing channel and the qubit AD channel in appropriate parameter regimes.

We first state the main convergence theorem. The assumptions below imply the strict concavity of the coherent information, and hence the uniqueness of the optimizer. 
Under the additional assumption that the optimizer is positive definite, these assumptions further yield the exponential convergence of the quantum Blahut-Arimoto algorithm.

\begin{theorem}\label{thm:BAexponential}
    Let $\mathcal{X}$ and $\mathcal{Y}$ be Hilbert spaces of dimensions $m$ and $n$, respectively. Let $\Phi \in \mathcal{T}(\mathcal{X},\mathcal{Y})$ be a degradable quantum channel with a degrading channel $\mathcal{D}$. Assume that $\Phi$ is injective and $\eta(\mathcal{D})<1$. Then $I_{\mathrm{coh}}(\rho , \Phi)$
    is strictly concave with respect to $\rho$. Consequently, the optimizer $\rho^{\star}$ of $I_{\mathrm{coh}}(\Phi)$ is unique.

    Assume additionally that $\rho^{\star}>0$. Let $\left\{\rho^{(t)}\right\}_{t\in\mathbb{N}^{+}}$ be the sequence generated by the quantum Blahut-Arimoto algorithm with parameter $\gamma\in[\zeta_{\mathrm{con}}(\Phi),1]$. Then there exist $N_{0}\in\mathbb{N}^{+}$ and a constant $a \in (0,\gamma]$ such that, for all $t \geq N_{0}$,
    \begin{align}
        I_{\mathrm{coh}}(\Phi)-I_{\mathrm{coh}}^{(t)}(\Phi)
        \leq
        \gamma
        \left(1-\frac{a}{\gamma}\right)^{t-N_{0}}
        D\left(\rho^{\star} \| \rho^{(N_{0})}\right).
    \end{align}
\end{theorem}

In Corollary~\ref{cor:examples} we verify the above assumptions for two important families of qubit channels. The qubit dephasing channel with parameter $p\in(0,1)$ is defined by
\begin{align}
    \mathcal{Z}_{p}(\rho)
    =
    (1-p)\rho+pZ\rho Z,
\end{align}
where $Z=\ket{0}\bra{0}-\ket{1}\bra{1}$ is the Pauli-$Z$ operator.

The qubit AD channel with parameter $p\in[0,1]$ is defined by
\begin{align}
    \mathcal{A}_{p}(\rho)=A_{0}\rho A_{0}^{\dagger}+A_{1}\rho A_{1}^{\dagger},
\end{align}
where $A_{0}=\ket{0}\bra{0}+\sqrt{1-p}\ket{1}\bra{1}$ and $A_{1}=\sqrt{p}\ket{0}\bra{1}$.
\begin{corollary}\label{cor:examples}
    The qubit dephasing channel $\mathcal{Z}_{p}$ with parameter $p\in(0,\frac{1}{2})\cup(\frac{1}{2},1)$ and the qubit AD channel $\mathcal{A}_{p}$ with parameter $p\in[0,\frac{1}{2})$ are injective and degradable. Moreover, the contraction coefficients of their respective degrading channels are strictly less than $1$, and the optimizers of $I_{\mathrm{coh}}(\mathcal{Z}_{p})$ and $I_{\mathrm{coh}}(\mathcal{A}_{p})$ are positive definite.
\end{corollary}

\subsection{Contraction coefficients for AD channels}\label{sec:AD}

In this subsection we first sharpen the existing estimate
\begin{align}
    \eta(\mathcal{A}_{p}) \leq \sqrt{1-p}
\end{align}
for the contraction coefficient of a single qubit AD channel~\cite[Lemma~6.3]{ref28}.
Second, we give explicit formulas for the relative contraction and expansion coefficients between two AD channels.
 
Our first theorem gives an exact one-dimensional variational formula for the contraction coefficient of $\mathcal{A}_{p}$.
The endpoint cases are trivial: $\eta(\mathcal{A}_{0})=1$ and $\eta(\mathcal{A}_{1})=0$.

\begin{theorem}\label{thm:contraction} 
    For $p\in(0,1)$, the contraction coefficient of $\mathcal{A}_{p}$ with respect to the quantum relative entropy is given by
    \begin{align}
        \eta(\mathcal{A}_{p})
        =
        \sup_{r\in[0,1)}
        (1-p)
        \frac{r}{\arctanh(r)}
        \frac{\arctanh\left(p+(1-p)r\right)}{p+(1-p)r} .
    \end{align}
    Here the expression at $r=0$ is understood by taking the limit.
\end{theorem}

Next, we consider the comparison between two AD channels. Let $0<p_{1}<p_{2}<1$ and $q=\frac{p_{2}-p_{1}}{1-p_{1}}$. By the composition rule for AD channels $\mathcal{A}_{p_{2}}=\mathcal{A}_{q} \circ \mathcal{A}_{p_{1}}$. The following theorem gives the corresponding relative contraction coefficient.

\begin{theorem}\label{thm:relcontraction}
    For $0<p_{1}<p_{2}<1$ and $q=\frac{p_{2}-p_{1}}{1-p_{1}}$, we have
    \begin{align}
    \eta\left(\mathcal{A}_{p_{2}},\mathcal{A}_{p_{1}}\right)
    =
    \sup_{r \in I_{p_{1}}}
    (1-q)\frac{r}{\arctanh(r)}
    \frac{\arctanh\left(q+(1-q)r\right)}{q+(1-q)r} ,
    \end{align}
    where
    $I_{p_{1}}=
    \begin{cases}
    [0,1), & p_{1}\in\left(0,\frac{1}{2}\right], \\
    [2p_{1}-1,1), & p_{1}\in\left(\frac{1}{2},1\right).
    \end{cases}$
    The expression at $r=0$ is understood by taking the limit.
\end{theorem}

Finally, we consider the reverse comparison, namely $\eta\left(\mathcal{A}_{p_{1}},\mathcal{A}_{p_{2}}\right)$ for $0<p_{1}<p_{2}<1$. This is equivalent to computing the relative expansion coefficient $\check{\eta}\left(\mathcal{A}_{p_{2}},\mathcal{A}_{p_{1}}\right)=1/\eta\left(\mathcal{A}_{p_{1}},\mathcal{A}_{p_{2}}\right)$. The following theorem gives its exact value.
\begin{theorem}\label{thm:relexpansion}
    For $0<p_{1}<p_{2}<1$, we have 
    \begin{align}
    \check{\eta}\left(\mathcal{A}_{p_{2}},\mathcal{A}_{p_{1}}\right)
    =\frac{p_{1}(1-p_{2})}{p_{2}(1-p_{1})} .\label{eqn:expansion}
    \end{align}
\end{theorem}

\section{Proof of the exponential convergence of the quantum Blahut-Arimoto algorithm}\label{sec:proof}

\subsection{Proof of Theorem~\ref{thm:BAexponential}}

\begin{proof}
Arbitrarily choose an ensemble $\left\{ p_{i},\rho_{i} \right\}$ and let $\rho=\sum\limits_{i}p_{i}\rho_{i}$. Then we have that
\begin{align}
    &\hspace{5mm}
    I_{\mathrm{coh}} \left(\sum_{i}p_{i}\rho_{i},\Phi \right)-\sum_{i}p_{i}I_{\mathrm{coh}}\left(\rho_{i},\Phi\right) \nonumber\\
    &= S \left( \Phi(\rho) \right) - S \left( \Phi^{c}(\rho) \right) - \sum_{i}p_{i} S \left( \Phi(\rho_{i}) \right) + \sum_{i}p_{i} S \left( \Phi^{c}(\rho_{i}) \right) \\
    &= \left[ S \left( \Phi(\rho) \right) - \sum_{i}p_{i} S \left( \Phi(\rho_{i}) \right) \right] - \left[ S \left( \Phi^{c}(\rho) \right) - \sum_{i}p_{i} S \left( \Phi^{c}(\rho_{i}) \right) \right] \\
    &= \sum_{i}p_{i}D \left( \Phi(\rho_{i}) \| \Phi(\rho) \right) - \sum_{i}p_{i}D \left( \Phi^{c}(\rho_{i}) \| \Phi^{c}(\rho) \right) \\
    &= \sum_{i}p_{i} \left[ D \left( \Phi(\rho_{i}) \| \Phi(\rho) \right) - D \left( \mathcal{D} \circ \Phi(\rho_{i}) \| \mathcal{D} \circ \Phi(\rho) \right) \right] \label{eqn:deg}\\
    &\geq \sum_{i}p_{i} \left( 1-\eta(\mathcal{D}) \right) D \left( \Phi(\rho_{i}) \| \Phi(\rho) \right)\label{eqn:sdpi},
\end{align}
where~\eqref{eqn:deg} follows from the assumption that $\Phi$ is degradable and~\eqref{eqn:sdpi} follows from the strong data processing inequality for the degrading channel $\mathcal{D}$. From the injectivity of $\Phi$ we have that $D \left( \Phi(\rho_{i}) \| \Phi(\rho) \right) > 0$.

Since $\eta(\mathcal{D})<1$, the above calculation shows that $I_{\mathrm{coh}}(\rho,\Phi)$ is strictly concave with respect to $\rho$, which implies that $\rho^{\star}$ is unique.
Next we show that $\rho^{(t)}\to\rho^{\star}$. Since $\gamma\geq\zeta_{\mathrm{con}}(\Phi)$, identity~\eqref{eq:Jidentity} gives
\begin{align}
    I_{\mathrm{coh}}^{(t)}(\Phi)=J_{\gamma}(\rho^{(t+1)},\rho^{(t)},\Phi)
    \leq J_{\gamma}(\rho^{(t+1)},\rho^{(t+1)},\Phi)
    =I_{\mathrm{coh}}(\rho^{(t+1)},\Phi)\leq I_{\mathrm{coh}}(\Phi),
\end{align}
and $I_{\mathrm{coh}}^{(t)}(\Phi)\to I_{\mathrm{coh}}(\Phi)$ by property~(1) in Section~\ref{sec:prelim}. Hence $I_{\mathrm{coh}}(\rho^{(t)},\Phi)\to I_{\mathrm{coh}}(\Phi)$. Since $\mathcal{D}(\mathcal{X})$ is compact and $\rho\mapsto I_{\mathrm{coh}}(\rho,\Phi)$ is continuous~\cite{ref25}, every limit point of $\{\rho^{(t)}\}$ is a maximizer of $I_{\mathrm{coh}}(\cdot,\Phi)$ and hence equals $\rho^{\star}$. Therefore $\|\rho^{(t)}-\rho^{\star}\|_{\infty}\to 0$ as $t\to+\infty$.

If an iterate equals $\rho^{\star}$, subsequent iterates remain at the optimizer and the desired conclusion is immediate. Otherwise let $\rho^{(t)}=\rho^{*}+r_{t}H_{t}$ where $r_{t}>0$ and $H_{t}$ is a traceless Hermitian operator with $\|H_{t}\|=1$.
Define
\begin{align}
    \chi_{t}(x)=\mathrm{Tr}\left[\left(\rho^{\star}\right)^{2}(\rho^{\star}+xH_{t})^{-1}\right]-1.
\end{align}
Let $S_{t}(x)=\left( \rho^{\star}+xH_{t} \right)^{-1}$. Then $S_{t}'(x)=-S_{t}(x)H_{t}S_{t}(x)$.
Hence we have
\begin{align}
    \chi_{t}'(x)=
    \mathrm{Tr} \left[ (\rho^{\star})^{2} S_{t}'(x) \right]
    = -\mathrm{Tr} \left[ (\rho^{\star})^{2} S_{t}(x)H_{t}S_{t}(x) \right] .
\end{align}
Further calculation gives
\begin{align}
    \chi_{t}''(x)
    = 2\mathrm{Tr} \left[ (\rho^{\star})^{2} S_{t}(x)H_{t}S_{t}(x)H_{t}S_{t}(x) \right]
\end{align}
and
\begin{align}
    \chi_{t}'''(x)
    = -6 \mathrm{Tr} \left[ (\rho^{\star})^{2} S_{t}(x)H_{t}S_{t}(x)H_{t}S_{t}(x)H_{t}S_{t}(x) \right] .
\end{align}
By Taylor's formula around $x=0$, using $\chi_{t}(0)=\mathrm{Tr}(\rho^{\star})-1=0$ and $\chi_{t}'(0)=-\mathrm{Tr}(H_{t})=0$, we have
\begin{align}
    \chi_{t}(r_{t})
    &=\chi_{t}(0)+r_{t}\chi_{t}'(0)+\frac{r_{t}^{2}}{2}\chi_{t}''(0)+\frac{r_{t}^{3}}{6}\chi_{t}'''(\xi_{t}) \\
    &=\frac{r_{t}^{2}}{2}\chi_{t}''(0)+\frac{r_{t}^{3}}{6}\chi_{t}'''(\xi_{t})
\end{align}
where $\xi_{t}\in(0,r_{t})$. By H\"older's inequality,
\begin{align}
    |\chi_{t}'''(\xi_{t})|
    &= 6 \left|\mathrm{Tr}\left[(\rho^{\star})^{2}S_{t}(\xi_{t})H_{t}S_{t}(\xi_{t})H_{t}S_{t}(\xi_{t})H_{t}S_{t}(\xi_{t})\right]\right| \\
    &\leq 6 \|(\rho^{\star})^{2}\|_{1} \|S_{t}(\xi_{t})\|_{\infty}^{4}\|H_{t}\|_{\infty}^{3}
    \leq 6 \|S_{t}(\xi_{t})\|_{\infty}^{4} .
\end{align}

We denote by $0<\lambda_{1} \leq \lambda_{2} \leq \dots \leq \lambda_{m}$
the eigenvalues of $\rho^{\star}$. It is obvious that $\langle \rho^{\star}-\rho^{(t)},(\rho^{\star})^{-1}(\rho^{\star}-\rho^{(t)}) \rangle \leq
\frac{1}{\lambda_{1}} \|\rho^{\star}-\rho^{(t)}\|_{2}^{2}$.

Fix a constant $0<\delta<\lambda_{1}$. Then there exists an integer $N' \in \mathbb{N}$ such that $r_{t}<\delta$ for all $t\geq N'$.
For the rest of this argument, fix $t\geq N'$. Since $\|H_{t}\|_{\infty}=1$ and $0<\xi_{t}<r_{t}<\delta$, we have $\rho^{\star}+\xi_{t} H_{t} \geq (\lambda_{1}-\delta)I_{\mathcal{X}}$.
It follows that $\|S_{t}(\xi_{t})\|_{\infty}=\|(\rho^{\star}+\xi_{t} H_{t})^{-1}\|_{\infty} \leq \frac{1}{\lambda_{1}-\delta}$.
Consequently, $|\chi_{t}'''(\xi_{t})|\leq C_{0}:=6(\lambda_{1}-\delta)^{-4}$, an upper bound independent of $t$.
 
Since $\chi_{t}''(0)=2\mathrm{Tr}\left(H_{t}(\rho^{\star})^{-1}H_{t}\right)$ and $\|\rho^{\star}-\rho^{(t)}\|_{2}\geq\|\rho^{\star}-\rho^{(t)}\|_{\infty}=r_{t}$, we obtain
\begin{align}
    \chi^{2}(\rho^{\star} \| \rho^{(t)})
    =\chi_{t}(r_{t})
    &= r_{t}^{2}\,\mathrm{Tr}\left(H_{t}(\rho^{\star})^{-1}H_{t}\right)+\frac{r_{t}^{3}}{6}\chi_{t}'''(\xi_{t}) \\
    &= \langle \rho^{\star}-\rho^{(t)},(\rho^{\star})^{-1}\left(\rho^{\star}-\rho^{(t)}\right) \rangle + \frac{r_{t}^{3}}{6}\chi_{t}'''(\xi_{t}) \\
    &\leq \left(\frac{1}{\lambda_{1}}+\frac{C_{0}r_{t}}{6}\right) \|\rho^{\star}-\rho^{(t)}\|_{2}^{2} .
\end{align}
Choosing $N_{0}\geq N'$ such that $C_{0}r_{t}/6\leq 1/\lambda_{1}$ for all $t\geq N_{0}$, we conclude that
\begin{align}
    \chi^{2}(\rho^{\star} \| \rho^{(t)}) \leq \frac{2}{\lambda_{1}} \|\rho^{\star}-\rho^{(t)}\|_{2}^{2}
    \qquad \text{for all } t \geq N_{0}. \label{eq:chibound}
\end{align}

For every $X \in \mathcal{L}(\mathcal{X})$, we have $\langle X,\Phi^{\dagger}\Phi(X) \rangle= \langle \Phi(X),\Phi(X) \rangle\geq 0$.
Moreover, equality holds if and only if $\Phi(X)=0$, which, by the injectivity of $\Phi$, is equivalent to $X=0$. Thus $\Phi^{\dagger}\Phi$ is positive definite. Therefore its smallest eigenvalue satisfies
$\mu_{\min}:=\lambda_{\min}(\Phi^{\dagger}\Phi)>0$.

In conclusion, we derive that 
\begin{align}
        \frac{D \left( \Phi\left(\rho^{\star}\right) \| \Phi\left(\rho^{(t)}\right) \right) - D \left( \Phi^{c}\left(\rho^{\star}\right) \| \Phi^{c}\left(\rho^{(t)}\right) \right)}{D\left(\rho^{\star} \| \rho^{(t)}\right)}
        & \geq \left( 1-\eta(\mathcal{D}) \right) \frac{D \left( \Phi\left(\rho^{\star}\right) \| \Phi\left(\rho^{(t)}\right) \right)}{D\left(\rho^{\star} \| \rho^{(t)}\right)} \label{eq:pa}\\
        & \geq \left( 1-\eta(\mathcal{D}) \right) \frac{D \left( \Phi\left(\rho^{\star}\right) \| \Phi\left(\rho^{(t)}\right) \right)}{\chi^{2}\left(\rho^{\star} \| \rho^{(t)}\right)} \label{eq:pb}\\
        & \geq \frac{\lambda_{1} \left(1-\eta(\mathcal{D})\right)}{4}\frac{\|\Phi\left(\rho^{\star}-\rho^{(t)}\right)\|_{2}^{2}}{\|\rho^{\star}-\rho^{(t)}\|_{2}^{2}} \label{eq:pc}\\
        & = \frac{\lambda_{1} \left(1-\eta(\mathcal{D})\right)}{4}\frac{\langle \rho^{\star}-\rho^{(t)} , \Phi^{\dagger}\Phi\left(\rho^{\star}-\rho^{(t)}\right) \rangle}{\|\rho^{\star}-\rho^{(t)}\|_{2}^{2}} \\
        & \geq \frac{\lambda_{1}\mu_{\min}\left(1-\eta(\mathcal{D})\right)}{4}
    \end{align}
for all $t \geq N_{0}$, where~\eqref{eq:pa} follows from the degradability of $\Phi$ and the definition of $\eta(\mathcal{D})$, \eqref{eq:pb} follows from~\eqref{eq:chi}, and~\eqref{eq:pc} follows from \eqref{eq:chibound} and the inequality $D(\rho \| \sigma)\geq \frac{1}{2}\mathrm{Tr}\left[ (\rho - \sigma)^{2} \right]$~\cite{ref40}, a consequence of the quantum Pinsker inequality $D(\rho\|\sigma)\geq\frac{1}{2}\|\rho-\sigma\|_{1}^{2}$ and $\|\cdot\|_{1}\geq\|\cdot\|_{2}$.

Let $a=\lambda_{1}\mu_{\min}\left(1-\eta(\mathcal{D})\right)/4$. Then the one-step estimate~\eqref{eqn:dp} holds for all $t \geq N_{0}$.
Iterating~\eqref{eq:ad} gives $D(\rho^{\star}\|\rho^{(t)})\leq\left(1-\frac{a}{\gamma}\right)^{t-N_{0}}D(\rho^{\star}\|\rho^{(N_{0})})$ for all $t\geq N_{0}$, and combining this with~\eqref{eq:ac} we obtain
\begin{align}
    I_{\mathrm{coh}}(\Phi)-I_{\mathrm{coh}}^{(t)}(\Phi) \leq \gamma \left(1-\frac{a}{\gamma}\right)^{t-N_{0}} D\left(\rho^{\star} \| \rho^{(N_{0})}\right)
\end{align}
for all $t \geq N_{0}$.
\end{proof}

\subsection{Proof of Corollary~\ref{cor:examples}}

\emph{Case 1: Qubit dephasing channel}

In Bloch coordinates, $\mathcal{Z}_{p}$ maps $(w_{1},w_{2},w_{3})$ to $((1-2p)w_{1},(1-2p)w_{2},w_{3})$. Hence $\mathcal{Z}_{p}$ is injective for $p\neq\frac{1}{2}$.
We focus on the remaining conditions.

A Kraus representation of $\mathcal{Z}_{p}$ is given by
\begin{align}
    \mathcal{Z}_{p}(\rho)=E_{0} \rho E_{0}^{\dagger}+E_{1} \rho E_{1}^{\dagger},
\end{align}
where
\begin{align}
    E_{0}=
    \begin{pmatrix}
        \sqrt{1-p} & 0 \\
        0 & \sqrt{1-p}
    \end{pmatrix},
    \quad
    E_{1}=
    \begin{pmatrix}
        \sqrt{p} & 0 \\
        0 & -\sqrt{p}
    \end{pmatrix}.
\end{align}
Let $V = E_{0} \otimes \ket{0} + E_{1} \otimes \ket{1}$
be the Stinespring isometry associated with this Kraus representation~\cite{ref41}.
Then the complementary channel is obtained by tracing out the output system, namely $\mathcal{Z}_{p}^{c}(\rho)=\mathrm{Tr}_{\mathcal{Y}}\left(V \rho V^{\dagger}\right)$.
A direct computation gives  
\begin{align}
    \mathcal{Z}_{p}^{c}(\rho)
    =
    \begin{pmatrix}
        (1-p)(\rho_{00}+\rho_{11})
        &
        \sqrt{p(1-p)}(\rho_{00}-\rho_{11})
        \\
        \sqrt{p(1-p)}(\rho_{00}-\rho_{11})
        &
        p(\rho_{00}+\rho_{11})
    \end{pmatrix}.
\end{align}
In particular, $\mathcal{Z}_{p}^{c}(\rho)$ depends only on the diagonal entries of $\rho$.
Since $\mathcal{Z}_{p}$ preserves the diagonal entries, we have $\mathcal{Z}_{p}^{c} \circ \mathcal{Z}_{p}=\mathcal{Z}_{p}^{c}$.
Thus the degrading channel may be chosen as $\mathcal{D}=\mathcal{Z}_{p}^{c}$, which implies that $\mathcal{Z}_{p}$ is degradable for all $p\in(0,1)$.

By the preceding analysis, the complementary channel of $\mathcal{Z}_{p}$ can be chosen as its degrading channel.
The proof of 
\begin{align}
    \eta(\mathcal{D})<1 \label{eq:ca}
\end{align}
is deferred to Appendix~\ref{ap:A}.

For any $P\in\{\sigma_x,\sigma_y,\sigma_z\}$, we have $\mathcal{Z}_p(P\rho P)=P\mathcal{Z}_p(\rho)P$ and hence $S(\mathcal{Z}_p(P\rho P))=S(\mathcal{Z}_p(\rho))$. 
Moreover, from the explicit expression of the complementary channel, we have 
\begin{align}
    \mathcal{Z}_p^c(P\rho P)
    =U_P\mathcal{Z}_p^c(\rho)U_P^\dagger,
\end{align}
where
\begin{align}
    U_P=
    \begin{cases}
    I_2, & P=\sigma_z,\\
    \sigma_z, & P=\sigma_x,\sigma_y.
    \end{cases}
\end{align}
Consequently, we obtain 
\begin{align}
    I_{\mathrm{coh}}(P\rho P,\mathcal{Z}_p)
    =I_{\mathrm{coh}}(\rho,\mathcal{Z}_p).
\end{align}
Since the Pauli twirl satisfies
\begin{align}
    \frac14\sum_{P\in\{I_2,\sigma_x,\sigma_y,\sigma_z\}}
P\rho P=\frac{I_2}{2},
\end{align}
the concavity of coherent information yields
\begin{align}
I_{\mathrm{coh}}\left(\frac{I_2}{2},\mathcal{Z}_p\right)
&=I_{\mathrm{coh}}\left(
\frac14\sum_P P\rho P,\mathcal{Z}_p\right)\\
&\geq\frac14\sum_P
I_{\mathrm{coh}}(P\rho P,\mathcal{Z}_p)\\
&=I_{\mathrm{coh}}(\rho,\mathcal{Z}_p).
\end{align}
Thus $I_2/2$ is a global optimizer. By the uniqueness
established in Theorem~\ref{thm:BAexponential},
we conclude that $\rho^\star=\frac{I_2}{2}>0$.

\noindent\emph{Case 2: Qubit AD channel}

In Bloch coordinates, $\mathcal{A}_{p}$ maps $(w_{1},w_{2},w_{3})$ to $(\sqrt{1-p}\,w_{1},\sqrt{1-p}\,w_{2},(1-p)w_{3}+p)$. Hence $\mathcal{A}_{p}$ is injective for $p<1$.

It is known that $\mathcal{A}_{p}$ is degradable for all $p\in[0,\frac{1}{2}]$~\cite{ref35,ref36}.
More precisely, for the Stinespring isometry $V=A_{0}\otimes\ket{0}+A_{1}\otimes\ket{1}$, one has $\mathcal{A}_{p}^{c}(\rho)=\sum\limits_{i,j}\mathrm{Tr}(A_{i}\rho A_{j}^{\dagger})\ket{i}\bra{j}$, and therefore
\begin{align}
    \mathcal{A}_{p}^{c}=\mathcal{A}_{1-p}=\mathcal{A}_{\frac{1-2p}{1-p}}\circ\mathcal{A}_{p}.
\end{align}
Thus, for $p\in[0,\frac{1}{2}]$, a degrading channel of $\mathcal{A}_{p}$ is given by $\mathcal{D}=\mathcal{A}_{\frac{1-2p}{1-p}}$.

It follows from $p\in[0,\frac{1}{2})$ that $\frac{1-2p}{1-p}\in(0,1]$. Therefore, to prove $\eta(\mathcal{D})<1$, it is enough to show that $\eta(\mathcal{A}_{p})<1$ for every $p\in(0,1]$. The case $p=1$ is trivial, since $\mathcal{A}_{1}$ is a constant channel and hence $\eta(\mathcal{A}_{1})=0$.
It remains to consider $p\in(0,1)$. By Theorem~\ref{thm:contraction}, we have
\begin{align}
    \eta(\mathcal{A}_{p})=
    \sup_{r\in[0,1)}
    (1-p)
    \frac{r}{\arctanh(r)}
    \frac{\arctanh\left(p+(1-p)r\right)}{p+(1-p)r}.
\end{align}
Let
\begin{align}
    \vartheta_{p}(r)
    =
    (1-p)
    \frac{r}{\arctanh(r)}
    \frac{\arctanh\left(p+(1-p)r\right)}{p+(1-p)r}. \label{eq:varthetadef}
\end{align}
The proof of 
\begin{align}
    \sup_{r\in[0,1)}\vartheta_{p}(r)<1\label{eq:cb}
\end{align}
is deferred to Appendix~\ref{ap:B}.

For $U_{\theta}=\mathrm{diag}(1,e^{i\theta})$, the AD channel satisfies $\mathcal{A}_{p}(U_{\theta}\rho U_{\theta}^{\dagger})
=U_{\theta}\mathcal{A}_{p}(\rho)U_{\theta}^{\dagger}$. Since $\mathcal{A}_{p}^{c}=\mathcal{A}_{1-p}$, the same covariance holds for the complementary channel. 
Define
\[
\overline{\rho}
=\frac{1}{2\pi}\int_{0}^{2\pi}
U_{\theta}\rho U_{\theta}^{\dagger}\,d\theta
=\mathrm{diag}(\rho_{00},\rho_{11}).
\]
By concavity,
\begin{align*}
I_{\mathrm{coh}}(\overline{\rho},\mathcal{A}_{p})
&\geq\frac{1}{2\pi}\int_{0}^{2\pi}
I_{\mathrm{coh}}(U_{\theta}\rho U_{\theta}^{\dagger},
\mathcal{A}_{p})\,d\theta\\
&=I_{\mathrm{coh}}(\rho,\mathcal{A}_{p}).
\end{align*}
By the uniqueness of the optimizer established in
Theorem~\ref{thm:BAexponential}, we conclude that
\[
\rho^{\star}=\mathrm{diag}(1-x^{\star},x^{\star}),
\]
where $x^{\star}$ maximizes
\[
\kappa_{p}(x)
:=I_{\mathrm{coh}}(\mathrm{diag}(1-x,x),\mathcal{A}_{p})
=h_{\mathrm{b}}((1-p)x)-h_{\mathrm{b}}(px),
\quad x\in[0,1],
\]
with $h_{\mathrm{b}}(y)=-y\log y-(1-y)\log(1-y)$.

For $p=0$, $\kappa_{0}(x)=h_{\mathrm{b}}(x)$
attains its unique maximum at $x=1/2$,
so $\rho^{\star}=I_{2}/2>0$.

For $0<p<1/2$, we have
\[
\kappa_{p}'(x)
=(1-p)\log\frac{1-(1-p)x}{(1-p)x}
-p\log\frac{1-px}{px}.
\]
In particular,
\[
\lim_{x\to0^{+}}\kappa_{p}'(x)
=\lim_{x\to0^{+}}
\left((1-2p)\log\frac{1}{x}+O(1)\right)
=+\infty,
\]
while
\[
\lim_{x\to1^{-}}\kappa_{p}'(x)
=\log\frac{p}{1-p}<0.
\]
Since $\kappa_{p}$ is continuous on $[0,1]$,
neither endpoint can be a maximizer.
Hence $x^{\star}\in(0,1)$, and therefore
\[
\rho^{\star}
=\mathrm{diag}(1-x^{\star},x^{\star})>0.
\]

\section{Proofs of the results on contraction coefficients of AD channels}\label{sec:contractionproof}

\subsection{Proof of Theorem~\ref{thm:contraction}} \label{sec:PTB}

We denote by $\Pxz:=\{\vec{u}\in\mathbb{R}^{3}:u_{2}=0\}$ the $xz$-plane and by $\Py:=\{\vec{u}\in\mathbb{R}^{3}:u_{1}=u_{3}=0\}$ the $y$-axis. By Lemma~\ref{lemma:BKM}, to prove Theorem~\ref{thm:contraction} it suffices to establish the identity
\begin{align}
    \eta_{\mathrm{BKM}}(\mathcal{A}_{p})
    &:=\sup_{\substack{\rho \in \mathcal{D}(\mathcal{X}) \\ X=X^{\dagger},\ \mathrm{Tr}(X)=0 \\ X \neq 0}}
    \frac{g_{\mathcal{A}_{p}(\rho)} \left( \mathcal{A}_{p}(X) \right)}{g_{\rho}(X)} \label{eqn:contraction}\\
    &=
    \sup_{r\in[0,1)}
    (1-p)\frac{r}{\arctanh(r)}\frac{\arctanh\left(p+(1-p)r\right)}{p+(1-p)r}. \label{eq:as}
 \end{align}

First, using the rotational covariance of the qubit AD channel around the $z$-axis, we show that the maximization in~\eqref{eqn:contraction} over all input states can be restricted to states whose Bloch vectors lie in $\Pxz$.
Then, using the special structure of the BKM metric induced by the AD channel, we show that the maximization in~\eqref{eqn:contraction} over all tangent vectors can be decomposed into two cases: the tangent vector lies in $\Py$, or it lies in $\Pxz$.
We then compute the maximum value when the states lie in $\Pxz$ and the tangent vectors lie in $\Py$, and show that this maximum value dominates the other case.
  
Throughout this subsection, we write 
\begin{align}
    \Theta(p)=\sup_{r\in[0,1)}\vartheta_{p}(r)=\sup_{r\in[0,1)}(1-p)\frac{r}{\arctanh(r)}\frac{\arctanh\left(p+(1-p)r\right)}{p+(1-p)r},
\end{align}
with $\vartheta_{p}$ as in~\eqref{eq:varthetadef}.

To prove the upper bound in~\eqref{eq:as}, it suffices to show $g_{\vec{w}_{\mathcal{A}_p}}(\vec{v}_{\mathcal{A}_p})\leq\Theta(p)g_{\vec{w}}(\vec{v})$ for all $|\vec{w}|<1$ and all $\vec{v}\in\mathbb{R}^{3}$. Sharpness will be established at the end of the proof.
If $|\vec{w}|=1$, namely if $\rho_{\vec{w}}$ is a pure state, then $g_{\vec{w}}(\vec{v})=+\infty$ for $\vec{v}\neq\vec{0}$, which implies the desired inequality is trivial.
Therefore, throughout the rest of this subsection, we assume that $|\vec{w}|<1$. Let
\begin{align}
    T_{\mathcal{A}_{p}}=\mathrm{diag}\left(\sqrt{1-p},\sqrt{1-p},1-p\right),
    \quad
    \vec{t}_{\mathcal{A}_{p}}= \left(0,0,p\right).
\end{align}
Then in Bloch coordinates, the AD channel $\mathcal{A}_{p}$ maps $\vec{w}=(w_{1},w_{2},w_{3})$ to
\begin{align}
    \vec{w}_{\mathcal{A}_{p}}
    &:= \left(\widetilde{w}_{1},\widetilde{w}_{2},\widetilde{w}_{3}\right) \\
    &= \left( \sqrt{1-p}\,w_{1}, \sqrt{1-p}\, w_{2},p+(1-p)w_{3} \right) \\
    &= T_{\mathcal{A}_{p}}\vec{w}+\vec{t}_{\mathcal{A}_{p}}
\end{align}
and maps $\vec{v}=(y_{1},y_{2},y_{3})$ to 
\begin{align}
    \vec{v}_{\mathcal{A}_{p}}
    &:= \left(\widetilde{y}_{1},\widetilde{y}_{2},\widetilde{y}_{3}\right) \\
    &= \left( \sqrt{1-p}\,y_{1}, \sqrt{1-p}\, y_{2},(1-p)y_{3} \right) \\
    &= T_{\mathcal{A}_{p}} \vec{v} .
\end{align}

Notice that the Bloch representation of $\mathcal{A}_{p}$ scales $w_{1}$ and $w_{2}$ by the same factor $\sqrt{1-p}$ and shifts only the $z$-component.
Hence $\mathcal{A}_{p}$ is covariant under rotations around the $z$-axis, namely applying a rotation around the $z$-axis to the input state before applying $\mathcal{A}_{p}$ gives the same result as first applying $\mathcal{A}_{p}$ and then applying the same rotation around the $z$-axis to the output state.
Moreover, for $p\in(0,1)$ and $|\vec{w}|<1$, a direct computation gives 
\begin{align}
    |\vec{w}_{\mathcal{A}_{p}}|<1 .\label{eq:cc}
\end{align}
Therefore, Lemma~\ref{lemma:formula} is applicable. The details are given in Appendix~\ref{ap:C}.

The following lemma indicates that we can restrict the optimization range of $\vec{w}$ to $\Pxz$ by using the rotational covariance of the qubit AD channel around the $z$-axis.
That is, we just need to prove $g_{\vec{w}_{\mathcal{A}_{p}}}(\vec{v}_{\mathcal{A}_{p}}) \leq \Theta(p) g_{\vec{w}}(\vec{v})$ for all $\vec{v}\in \mathbb{R}^{3}$ and all $\vec{w} \in \Pxz$ with $|\vec{w}|<1$.

\begin{lemma}
    Let $p\in(0,1)$ and $C(p)\geq0$.
    Suppose that $g_{\vec{w}_{\mathcal{A}_{p}}}(\vec{v}_{\mathcal{A}_{p}}) \leq C(p)\, g_{\vec{w}}(\vec{v})$ for all $\vec{v}\in \mathbb{R}^{3}$ and all $\vec{w} \in \Pxz$ with $|\vec{w}|<1$.
    Then $g_{\vec{w}_{\mathcal{A}_{p}}}(\vec{v}_{\mathcal{A}_{p}}) \leq C(p)\, g_{\vec{w}}(\vec{v})$ holds for all $\vec{v}\in \mathbb{R}^{3}$ and all $\vec{w} \in \mathbb{R}^{3}$ with $|\vec{w}| < 1$.\label{le:cov}
\end{lemma} 

\begin{proof}

Let $\varphi\in(-\pi,\pi]$ be the polar angle of the point $(w_{1},w_{2})\in\mathbb{R}^{2}$, i.e.,
\begin{align}
    (w_{1},w_{2})=\sqrt{w_{1}^{2}+w_{2}^{2}}\,(\cos\varphi,\sin\varphi)
\end{align}
with the convention $\varphi=0$ when $w_1=w_2=0$. Let

\begin{align}
    U=\begin{pmatrix}e^{\frac{i\varphi}{2}} & 0 \\ 0 & e^{-\frac{i\varphi}{2}}\end{pmatrix} .
\end{align}
It can be verified that $U$ is a unitary matrix and $U \rho_{\vec{w}} U^{\dagger}=\frac{1}{2}(I_{2}+\vec{w}_{U} \cdot \vec{\sigma})$
where 
\begin{align}
    \vec{w}_{U}=\left(\sqrt{w_{1}^{2}+w_{2}^{2}},\,0,\,w_{3}\right) \in \Pxz,
    \qquad |\vec{w}_{U}|=|\vec{w}|<1 .
\end{align}
Moreover, $UX_{\vec{v}}U^{\dagger}$ is again traceless and Hermitian, i.e., $UX_{\vec{v}}U^{\dagger}=X_{\vec{v}'}$ for some $\vec{v}'\in\mathbb{R}^{3}$.

Using the unitarity of $U$ we have that
\begin{align}
    (U\rho_{\vec{w}} U^{\dagger}+rI)^{-1}=\left(U(\rho_{\vec{w}}+rI) U^{\dagger}\right)^{-1}
    &=U(\rho_{\vec{w}}+rI)^{-1}U^{\dagger},
\end{align}
which, together with the definition of the BKM metric, implies that 
\begin{align}
    g_{\rho_{\vec{w}}}(X_{\vec{v}})
    =g_{U \rho_{\vec{w}} U^{\dagger}}(U X_{\vec{v}} U^{\dagger}).
\end{align}

Notice that $\mathcal{A}_{p}(U X U^{\dagger})=U \mathcal{A}_{p}(X) U^{\dagger}$ for all $X \in \mathcal{L}(\mathcal{X})$.
Then by direct calculations, we obtain   
\begin{align}
    g_{\vec{w}_{\mathcal{A}_{p}}}\left(\vec{v}_{\mathcal{A}_{p}}\right)
    &= g_{\mathcal{A}_{p}(\rho_{\vec{w}})}\left(\mathcal{A}_{p}(X_{\vec{v}})\right) \\
    &= g_{U \mathcal{A}_{p}(\rho_{\vec{w}}) U^{\dagger}}\left(U \mathcal{A}_{p}(X_{\vec{v}}) U^{\dagger}\right) \\
    &= g_{\mathcal{A}_{p}(U \rho_{\vec{w}} U^{\dagger})}\left(\mathcal{A}_{p}(U X_{\vec{v}} U^{\dagger})\right) \\
    &\leq C(p)\, g_{U \rho_{\vec{w}} U^{\dagger}}(U X_{\vec{v}} U^{\dagger}) \\
    &= C(p)\, g_{\rho_{\vec{w}}}(X_{\vec{v}}) \\
    &= C(p)\, g_{\vec{w}}(\vec{v})
\end{align}
for all $\vec{v}\in \mathbb{R}^{3}$ and all $\vec{w} \in \mathbb{R}^{3}$ with $|\vec{w}| < 1$.
\end{proof}

Using the formulas in Lemma~\ref{lemma:formula}, for $\vec{w}=(w_{1},0,w_{3})\in\Pxz$ we have the following decomposition:
\begin{align}
    g_{\vec{w}}(\vec{v})
    &= \frac{4|\vec{v}|^{2}}{1-|\vec{w}|^{2}}\left[\frac{|\vec{w}\cdot \vec{v}|^{2}}{|\vec{w}|^{2}|\vec{v}|^{2}}\left(1-f(|\vec{w}|)\right)+f(|\vec{w}|)\right] \\
    &= \frac{4}{1-|\vec{w}|^{2}}\left[\frac{1-f(|\vec{w}|)}{|\vec{w}|^{2}}(w_{1}y_{1}+w_{3}y_{3})^{2}+\left(y_{1}^{2}+y_{2}^{2}+y_{3}^{2}\right)f(|\vec{w}|)\right] \\
    &= g_{\vec{w}}(\vec{v}_{y})+g_{\vec{w}}(\vec{v}_{xz})
\end{align}
where $\vec{v}_{y}=(0,y_{2},0) \in \Py$ and $\vec{v}_{xz} = (y_{1},0,y_{3}) \in \Pxz$.

Similarly, we can decompose $g_{\vec{w}_{\mathcal{A}_{p}}}(\vec{v}_{\mathcal{A}_{p}})$ as follows:
\begin{align}
    g_{\vec{w}_{\mathcal{A}_{p}}}(\vec{v}_{\mathcal{A}_{p}})
    &= \frac{4|\vec{v}_{\mathcal{A}_{p}}|^{2}}{1-|\vec{w}_{\mathcal{A}_{p}}|^{2}}\left[\frac{|\vec{w}_{\mathcal{A}_{p}}\cdot \vec{v}_{\mathcal{A}_{p}}|^{2}}{|\vec{w}_{\mathcal{A}_{p}}|^{2}|\vec{v}_{\mathcal{A}_{p}}|^{2}}\left(1-f(|\vec{w}_{\mathcal{A}_{p}}|)\right)+f(|\vec{w}_{\mathcal{A}_{p}}|)\right] \\
    &= \frac{4}{1-|\vec{w}_{\mathcal{A}_{p}}|^{2}}\left[\frac{1-f(|\vec{w}_{\mathcal{A}_{p}}|)}{|\vec{w}_{\mathcal{A}_{p}}|^{2}}(\widetilde{w}_{1}\widetilde{y}_{1}+\widetilde{w}_{3}\widetilde{y}_{3})^{2}+\left(\widetilde{y}_{1}^{2}+\widetilde{y}_{2}^{2}+\widetilde{y}_{3}^{2}\right)f(|\vec{w}_{\mathcal{A}_{p}}|)\right] \\
    &= g_{\vec{w}_{\mathcal{A}_{p}}}\left(\left(\vec{v}_{\mathcal{A}_{p}}\right)_{y}\right)+g_{\vec{w}_{\mathcal{A}_{p}}}\left(\left(\vec{v}_{\mathcal{A}_{p}}\right)_{xz}\right) ,
\end{align}
where $\left(\vec{v}_{\mathcal{A}_{p}}\right)_{y}=(0,\widetilde{y}_{2},0) \in \Py$ and $\left(\vec{v}_{\mathcal{A}_{p}}\right)_{xz} = (\widetilde{y}_{1},0,\widetilde{y}_{3}) \in \Pxz$.

The above decomposition shows that the problem splits into two parts: $\vec{v}$ in $\Py$ and $\vec{v}$ in $\Pxz$.
These two cases are treated in Lemma~\ref{lem:yaxis} and Lemma~\ref{lem:xoz}, respectively.

\begin{lemma}\label{lem:yaxis}
    For $\vec{w} \in \Pxz$ with $|\vec{w}| < 1$ and $\vec{v} \in \Py$, one obtains $g_{\vec{w}_{\mathcal{A}_{p}}}\left(\vec{v}_{\mathcal{A}_{p}}\right) \leq \Theta(p)\, g_{\vec{w}}\left(\vec{v}\right)$.

\end{lemma}

\begin{proof}
Let
\begin{align}
    \vec{w}=\left(w_{1},0,w_{3}\right)=
    \left(r\cos\theta,0,r\sin\theta\right), \quad r\in[0,1), \quad \theta\in[0,2\pi),
\end{align}
and let $\vec{v}=\left(0,y_{2},0\right)$ for $y_{2}\in\mathbb{R}\backslash \{0\}.$

Let $m_{r}\left(\theta\right)\triangleq|\vec{w}_{\mathcal{A}_{p}}|$.
 A direct computation gives
\begin{align}
    m_{r}\left(\theta\right)
    &=\sqrt{\left(1-p\right)r^{2}\cos^{2}\theta+\left(p+\left(1-p\right)r\sin\theta\right)^{2}} \\
    &=\sqrt{\left(1-p\right)r^{2}\cos^{2}\theta+p^{2}+2p\left(1-p\right)r\sin\theta+\left(1-p\right)^{2}r^{2}\sin^{2}\theta}.
\end{align}
Since
\begin{align}
    m_{r}^{2}\left(\theta\right)
    =
    p^{2}+\left(1-p\right)r^{2}+2p\left(1-p\right)r\sin\theta-p\left(1-p\right)r^{2}\sin^{2}\theta,
\end{align}
$m_{r}^{2}\left(\theta\right)$ is a quadratic function of $\sin\theta$.
For fixed $r\in[0,1)$, it attains its maximum at $\sin\theta=1$.
Hence,
\begin{align}
    \max_{\theta\in[0,2\pi)}
    m_{r}\left(\theta\right)
    =m_{r}\left(\frac{\pi}{2}\right)
    =p+\left(1-p\right)r.
\end{align}
    
A direct calculation gives $\vec{w} \cdot \vec{v}=0$ and
$\vec{w}_{\mathcal{A}_{p}} \cdot \vec{v}_{\mathcal{A}_{p}}=0$.
Then by Lemma~\ref{lemma:formula} and the fact that the function $x\longmapsto\frac{\arctanh\left(x\right)}{x}$ is increasing on $[0,1)$ (see~\eqref{eq:fseries}), we have
\begin{align*}
    \ell_{y}(p)\triangleq
    &\sup\limits_{\substack{ \vec{w} \in \Pxz,\ |\vec{w}| < 1 \\ \vec{v} \in \Py,\ \vec{v}\neq \vec{0}}}
    \frac{g_{\vec{w}_{\mathcal{A}_{p}}}\left(\vec{v}_{\mathcal{A}_{p}}\right)}{g_{\vec{w}}\left(\vec{v}\right)} \\
    &= \sup\limits_{\substack{\vec{w} \in \Pxz,\ |\vec{w}| < 1}}\left(1-p\right)
    \frac{1-|\vec{w}|^{2}}{1-|\vec{w}_{\mathcal{A}_{p}}|^{2}}
    \frac{f\left(|\vec{w}_{\mathcal{A}_{p}}|\right)}{f\left(|\vec{w}|\right)} \\
    &=\sup\limits_{\substack{\vec{w} \in \Pxz,\ |\vec{w}| < 1}}\left(1-p\right)
    \frac{|\vec{w}|}{\arctanh\left(|\vec{w}|\right)}
    \frac{\arctanh\left(|\vec{w}_{\mathcal{A}_{p}}|\right)}
    {|\vec{w}_{\mathcal{A}_{p}}|} \\
    &=\sup\limits_{\substack{r\in[0,1) \\ \theta\in[0,2\pi)}}
    \left(1-p\right)
    \frac{r}{\arctanh\left(r\right)}
    \frac{\arctanh\left(m_{r}\left(\theta\right)\right)}
    {m_{r}\left(\theta\right)} \\
    &=\sup\limits_{r\in[0,1)}
    \left(1-p\right)
    \frac{r}{\arctanh\left(r\right)}
    \frac{\arctanh\left(p+\left(1-p\right)r\right)}
    {p+\left(1-p\right)r}\\
    &=\Theta(p).
\end{align*}

This completes the proof.
\end{proof}

\begin{lemma}\label{lem:xoz}
    For $\vec{w} \in \Pxz$ with $|\vec{w}| < 1$ and $\vec{v} \in \Pxz$, one obtains $g_{\vec{w}_{\mathcal{A}_{p}}}\left(\vec{v}_{\mathcal{A}_{p}}\right) \leq \Theta(p)\, g_{\vec{w}}\left(\vec{v}\right)$.
\end{lemma}

\begin{proof}

Let $\bar{p}:=1-p$ and
\begin{align}
    M
    &=\frac{1}{1-|\vec{w}|^{2}}
    \left[f\left(|\vec{w}|\right)I+\frac{1-f\left(|\vec{w}|\right)}{|\vec{w}|^{2}}\vec{w}\vec{w}^{T}\right] \\
    &=
    \frac{1}{1-|\vec{w}|^{2}}
    \begin{pmatrix}
    \dfrac{f\left(|\vec{w}|\right)w_{3}^{2}+w_{1}^{2}}{|\vec{w}|^{2}} & 0 
    & \dfrac{\left(1-f\left(|\vec{w}|\right)\right)w_{1}w_{3}}{|\vec{w}|^{2}} \\ 0 & f\left(|\vec{w}|\right)& 0 \\
    \dfrac{\left(1-f\left(|\vec{w}|\right)\right)w_{1}w_{3}}{|\vec{w}|^{2}}
    & 0 & \dfrac{f\left(|\vec{w}|\right)w_{1}^{2}+w_{3}^{2}}{|\vec{w}|^{2}}
    \end{pmatrix} \label{eq:ce}
\end{align}
and
\begin{align}
    N
    &=\frac{1}{1-|\vec{w}_{\mathcal{A}_{p}}|^{2}}T_{\mathcal{A}_{p}}^{T}
    \left[f\left(|\vec{w}_{\mathcal{A}_{p}}|\right)I+\frac{1-f\left(|\vec{w}_{\mathcal{A}_{p}}|\right)}{|\vec{w}_{\mathcal{A}_{p}}|^{2}}\vec{w}_{\mathcal{A}_{p}}\vec{w}_{\mathcal{A}_{p}}^{T}
    \right] T_{\mathcal{A}_{p}} \\
    &=\frac{1}{1-|\vec{w}_{\mathcal{A}_{p}}|^{2}}
    \begin{pmatrix}
    \bar{p} \,\dfrac{f\left(|\vec{w}_{\mathcal{A}_{p}}|\right)\widetilde{w}_{3}^{2}+\widetilde{w}_{1}^{2}}{|\vec{w}_{\mathcal{A}_{p}}|^{2}} & 0 & \bar{p}^{\frac{3}{2}}
    \dfrac{\left(1-f\left(|\vec{w}_{\mathcal{A}_{p}}|\right)\right)\widetilde{w}_{1}\widetilde{w}_{3}}{|\vec{w}_{\mathcal{A}_{p}}|^{2}} \\ 0 & \bar{p} \,f\left(|\vec{w}_{\mathcal{A}_{p}}|\right) & 0 \\
    \bar{p}^{\frac{3}{2}}
    \dfrac{\left(1-f\left(|\vec{w}_{\mathcal{A}_{p}}|\right)\right)\widetilde{w}_{1}\widetilde{w}_{3}}{|\vec{w}_{\mathcal{A}_{p}}|^{2}} & 0 & \bar{p}^{2}
    \dfrac{f\left(|\vec{w}_{\mathcal{A}_{p}}|\right)\widetilde{w}_{1}^{2}+\widetilde{w}_{3}^{2}}{|\vec{w}_{\mathcal{A}_{p}}|^{2}}
    \end{pmatrix}, \label{eq:cf}
\end{align}
so that $g_{\vec{w}}(\vec{v})=4\vec{v}^{T}M\vec{v}$ and $g_{\vec{w}_{\mathcal{A}_{p}}}(\vec{v}_{\mathcal{A}_{p}})=4\vec{v}^{T}N\vec{v}$.
The matrix formulas are interpreted by continuity when $\vec{w}=0$ or $\vec{w}_{\mathcal{A}_p}=0$. 

As in the proof of Lemma~\ref{lem:yaxis}, let
\begin{align}
    \ell_{xz}(p)
    &\triangleq \sup\limits_{\substack{\vec{w} \in \Pxz,\ |\vec{w}| < 1 \\ \vec{v} \in \Pxz,\ \vec{v} \neq \vec{0}}}
    \frac{g_{\vec{w}_{\mathcal{A}_{p}}}\left(\vec{v}_{\mathcal{A}_{p}}\right)}
    {g_{\vec{w}}\left(\vec{v}\right)} \\
    &= \sup\limits_{\substack{\vec{w} \in \Pxz,\ |\vec{w}| < 1 \\
    \vec{v} \in \Pxz,\ \vec{v} \neq \vec{0}}}
    \frac{\vec{v}^{T}N\vec{v}}{\vec{v}^{T}M\vec{v}} \\
    &\leq \sup\limits_{\substack{\vec{w} \in \Pxz,\ |\vec{w}| < 1}}
    \lambda_{\max}\left(M^{-\frac{1}{2}}NM^{-\frac{1}{2}}\right) \\
    &= \sup\limits_{\substack{\vec{w} \in \Pxz,\ |\vec{w}| < 1}}
    \lambda_{\max}\left(M^{-1}N\right).
\end{align}
The eigenvalues of $M^{-1}N$ can be characterized by
\begin{align}
    \lambda\in\sigma\left(M^{-1}N\right)
    \quad\Longleftrightarrow\quad
    \det\left(\lambda M-N\right)=0.
\end{align}
From the block structure of $M$ and $N$, the eigenvalues of $M^{-1}N$ consist of one eigenvalue $\lambda_{y}$ corresponding to $\Py$ and two eigenvalues $\lambda_{\pm}$ corresponding to $\Pxz$.
 A direct calculation gives
\begin{align}
    \lambda_{y}=
    \left(1-p\right)
    \frac{1-|\vec{w}|^{2}}{1-|\vec{w}_{\mathcal{A}_{p}}|^{2}}
    \frac{f\left(|\vec{w}_{\mathcal{A}_{p}}|\right)}{f\left(|\vec{w}|\right)}. \label{eq:lambday}
\end{align}
It can be shown that
\begin{align}
    \lambda_{\pm}\leq\lambda_{y}.\label{eq:cd}
\end{align}

The details of this spectral comparison are deferred to Appendix~\ref{ap:D}.
Consequently, by Lemma~\ref{lem:yaxis}, we have 
\begin{align}
    \ell_{xz}(p)
    \leq
    \sup\limits_{\substack{\vec{w} \in \Pxz,\ |\vec{w}| < 1}}
    \left(1-p\right)\frac{1-|\vec{w}|^{2}}{1-|\vec{w}_{\mathcal{A}_{p}}|^{2}}
    \frac{f\left(|\vec{w}_{\mathcal{A}_{p}}|\right)}{f\left(|\vec{w}|\right)}\notag
    \leq
    \Theta(p).
\end{align}

This completes the proof.
\end{proof}

Combining the above two lemmas, we have
\begin{align}
    \sup\limits_{\substack{\vec{w} \in \Pxz,\ |\vec{w}| < 1 \\ \vec{v}\in \mathbb{R}^{3},\ \vec{v}\neq \vec{0}}}\frac{g_{\vec{w}_{\mathcal{A}_{p}}}(\vec{v}_{\mathcal{A}_{p}})}{g_{\vec{w}}(\vec{v})}
    &=\sup\limits_{\substack{\vec{w} \in \Pxz,\ |\vec{w}| < 1 \\ \vec{v}\in \mathbb{R}^{3},\ \vec{v}\neq \vec{0}}}\frac{g_{\vec{w}_{\mathcal{A}_{p}}}\left(\left(\vec{v}_{\mathcal{A}_{p}}\right)_{y}\right)+g_{\vec{w}_{\mathcal{A}_{p}}}\left(\left(\vec{v}_{\mathcal{A}_{p}}\right)_{xz}\right)}{g_{\vec{w}}(\vec{v}_{y})+g_{\vec{w}}(\vec{v}_{xz})} \\
    &\leq \max \left\{ \sup\limits_{\substack{ \vec{w} \in \Pxz,\ |\vec{w}| < 1 \\ \vec{v} \in \Py,\ \vec{v}\neq \vec{0}}}
    \frac{g_{\vec{w}_{\mathcal{A}_{p}}}\left(\vec{v}_{\mathcal{A}_{p}}\right)}{g_{\vec{w}}\left(\vec{v}\right)} , \sup\limits_{\substack{\vec{w} \in \Pxz,\ |\vec{w}| < 1 \\ \vec{v} \in \Pxz,\ \vec{v} \neq \vec{0}}}
    \frac{g_{\vec{w}_{\mathcal{A}_{p}}}\left(\vec{v}_{\mathcal{A}_{p}}\right)}
    {g_{\vec{w}}\left(\vec{v}\right)} \right\} \\
    &\leq \Theta(p) .
\end{align}
Furthermore, it follows from the proof of Lemma~\ref{lem:yaxis} that the constant $\Theta(p)$ is sharp: the supremum $\Theta(p)$ is approached by taking $\vec{w}=(0,0,r)$ on the $z$-axis and $\vec{v}$ in $\Py$, for which the ratio equals $\vartheta_{p}(r)$. Together with Lemma~\ref{le:cov}, this proves~\eqref{eq:as}.

\subsection{Proof of Theorem~\ref{thm:relcontraction}}
For $p_{2}>p_{1}$, the channel $\mathcal{A}_{p_{2}}$ can be decomposed as
\begin{align}
    \mathcal{A}_{p_{2}}=\mathcal{A}_{q}\circ \mathcal{A}_{p_{1}},
    \quad q=\frac{p_{2}-p_{1}}{1-p_{1}} .
\end{align}
Therefore, the comparison between $\mathcal{A}_{p_{1}}$ and $\mathcal{A}_{p_{2}}$ is equivalent to studying the contraction induced by the additional AD channel $\mathcal{A}_{q}$ on the image of $\mathcal{A}_{p_{1}}$ instead of the whole Bloch ball.

By Lemma~\ref{lemma:BKM}, to prove Theorem~\ref{thm:relcontraction} it suffices to establish the identity
\begin{align}
    \eta_{\mathrm{BKM}}(\mathcal{A}_{p_{2}},\mathcal{A}_{p_{1}})
    &:=
    \sup_{\substack{\rho \in \mathcal{D}(\mathcal{X}) \\ X=X^{\dagger},\ \mathrm{Tr}(X)=0 \\ X \neq 0}}
    \frac{g_{\mathcal{A}_{p_{2}}(\rho)}\left(\mathcal{A}_{p_{2}}(X)\right)}
    {g_{\mathcal{A}_{p_{1}}(\rho)}\left(\mathcal{A}_{p_{1}}(X)\right)} \\
    &= \sup_{r \in I_{p_{1}}}
    (1-q)\frac{r}{\arctanh(r)}
    \frac{\arctanh\left(q+(1-q)r\right)}{q+(1-q)r} .
\end{align}

Let $\omega=\mathcal{A}_{p_{1}}(\rho)$ and $Y=\mathcal{A}_{p_{1}}(X)$.
Since $p_{1}\in(0,1)$, the restriction of $\mathcal{A}_{p_{1}}$ to the space of traceless Hermitian operators is a linear isomorphism onto itself.
Thus the problem is equivalent to proving
\begin{align}
    \sup_{\substack{\omega \in \mathcal{A}_{p_{1}}\left(\mathcal{D}(\mathcal{X})\right) \\ Y=Y^{\dagger},\ \mathrm{Tr}(Y)=0 \\ Y \neq 0}}
    \frac{g_{\mathcal{A}_{q}(\omega)}(\mathcal{A}_{q}(Y))}{g_{\omega}(Y)}
    =
    \sup_{r \in I_{p_{1}}}
    (1-q)\frac{r}{\arctanh(r)}
    \frac{\arctanh\left(q+(1-q)r\right)}{q+(1-q)r} .
\end{align}

For $0<p_1<1$, the only rank-deficient state
in the image of $\mathcal{A}_{p_1}$ is
$\ket{0}\bra{0}$, which is excluded from
the BKM optimization.
For any $\omega=\mathcal{A}_{p_1}(\rho)>0$,
let
\begin{align}
    \rho_\varepsilon
=(1-\varepsilon)\rho+\varepsilon I_2/2.
\end{align}
Then $\rho_\varepsilon>0$ and
$\mathcal{A}_{p_1}(\rho_\varepsilon)\to\omega$.
Since $\mathcal{A}_{p_1}$ is invertible on
traceless Hermitian operators and the BKM
ratio is continuous at positive definite
states, every ratio at $(\omega,Y)$ is
approximated by ratios arising from
positive definite inputs.
Hence enlarging the optimization domain
to the image ellipsoid with
$\ket{0}\bra{0}$ excluded changes neither
the supremum nor the infimum.

We now pass to Bloch coordinates. Write the Bloch vector of $\omega$ as $\vec{w}$ and the Pauli-coordinate vector of $Y$ as $\vec{v}$. 
The image $\mathcal{A}_{p_{1}}\left(\mathcal{D}(\mathcal{X})\right)$ is the ellipsoid
\begin{align}
    \mathbb{E}_{p_{1}}=\left\{\vec{w}=(w_{1},w_{2},w_{3})\in\mathbb{R}^{3}:\frac{w_{1}^{2}+w_{2}^{2}}{1-p_{1}}+\frac{(w_{3}-p_{1})^{2}}{(1-p_{1})^{2}}\leq 1\right\}.
\end{align}
As explained above, the rank-deficient output $\ket{0}\bra{0}$ is excluded from the BKM ratio, and it therefore suffices to consider $|\vec{w}|<1$.

By the rotational covariance of the qubit AD channel around the $z$-axis, as in Lemma~\ref{le:cov}, it suffices to restrict the Bloch vector $\vec{w}$ to $\Pxz$.
Therefore, we only need to consider $\vec{w}=(w_{1},0,w_{3})\in \mathbb{E}_{p_{1}}$, a condition which is equivalent to $\frac{w_{1}^{2}}{1-p_{1}}+\frac{(w_{3}-p_{1})^{2}}{(1-p_{1})^{2}}\leq 1$.
Consequently, it is sufficient to prove 
\begin{align}
    \sup_{\substack{\vec{w} \in \mathbb{E}^{xz}_{p_{1}},\ |\vec{w}|<1 \\
    \vec{v}\in\mathbb{R}^{3},\ \vec{v} \neq \vec{0}}}
    \frac{g_{\vec{w}_{\mathcal{A}_{q}}}\left(\vec{v}_{\mathcal{A}_{q}}\right)}
    {g_{\vec{w}}(\vec{v})}
    =
    \sup_{r \in I_{p_{1}}}
    (1-q)\frac{r}{\arctanh(r)}
    \frac{\arctanh\left(q+(1-q)r\right)}{q+(1-q)r} ,
\end{align}
where $\mathbb{E}^{xz}_{p_{1}}:=\mathbb{E}_{p_{1}}\cap\Pxz=\left\{ \vec{w}=(w_{1},0,w_{3})\in\mathbb{R}^{3}:\frac{w_{1}^{2}}{1-p_{1}}+\frac{(w_{3}-p_{1})^{2}}{(1-p_{1})^{2}}\leq 1 \right\}$.
 
As in the proof of Theorem~\ref{thm:contraction}, we can decompose $g_{\vec{w}}(\vec{v})$ and $g_{\vec{w}_{\mathcal{A}_{q}}}(\vec{v}_{\mathcal{A}_{q}})$ into two parts, respectively.
When $\vec{v}$ lies in $\Py$, we derive that
\begin{align}
    \sup_{\substack{\vec{w} \in \mathbb{E}^{xz}_{p_{1}},\ |\vec{w}|<1 \\ \vec{v} \in \Py,\ \vec{v}\neq\vec{0}}}
    \frac{g_{\vec{w}_{\mathcal{A}_{q}}}(\vec{v}_{\mathcal{A}_{q}})}{g_{\vec{w}}(\vec{v})}
    &= \sup_{\substack{\vec{w} \in \mathbb{E}^{xz}_{p_{1}},\ |\vec{w}|<1}}
    (1-q)
    \frac{1-|\vec{w}|^{2}}{1-|\vec{w}_{\mathcal{A}_{q}}|^{2}}
    \frac{f(|\vec{w}_{\mathcal{A}_{q}}|)}{f(|\vec{w}|)} .
\end{align}
When $\vec{v}$ lies in $\Pxz$, we have

\begin{align}
    \sup_{\substack{\vec{w} \in \mathbb{E}^{xz}_{p_{1}},\ |\vec{w}|<1 \\ \vec{v} \in \Pxz,\ \vec{v}\neq\vec{0}}}
    \frac{g_{\vec{w}_{\mathcal{A}_{q}}}(\vec{v}_{\mathcal{A}_{q}})}{g_{\vec{w}}(\vec{v})}
    &\leq \sup_{\substack{\vec{w} \in \mathbb{E}^{xz}_{p_{1}},\ |\vec{w}|<1}} \lambda_{\max}(M^{-1}N) \\
    &\leq \sup_{\substack{\vec{w} \in \mathbb{E}^{xz}_{p_{1}},\ |\vec{w}|<1}}
    (1-q)
    \frac{1-|\vec{w}|^{2}}{1-|\vec{w}_{\mathcal{A}_{q}}|^{2}}
    \frac{f(|\vec{w}_{\mathcal{A}_{q}}|)}{f(|\vec{w}|)} ,
\end{align}
where $M$ and $N$ are the matrices defined in~\eqref{eq:ce} and~\eqref{eq:cf} with $p$ replaced by $q$, and the second inequality is~\eqref{eq:cd} (Appendix~\ref{ap:D} applies to every $\vec{w}\in\Pxz$ with $|\vec{w}|<1$, with $p$ replaced by $q$).

In conclusion, 
\begin{align}
    \Pi(q)
    &\triangleq \sup_{\substack{\vec{w} \in \mathbb{E}^{xz}_{p_{1}},\ |\vec{w}|<1 \\
    \vec{v}\in\mathbb{R}^{3},\ \vec{v} \neq \vec{0}}}
    \frac{g_{\vec{w}_{\mathcal{A}_{q}}}\left(\vec{v}_{\mathcal{A}_{q}}\right)}
    {g_{\vec{w}}(\vec{v})} \\
    &= \sup_{\substack{\vec{w} \in \mathbb{E}^{xz}_{p_{1}},\ |\vec{w}|<1}}
    (1-q)
    \frac{1-|\vec{w}|^{2}}{1-|\vec{w}_{\mathcal{A}_{q}}|^{2}}
    \frac{f\left(|\vec{w}_{\mathcal{A}_{q}}|\right)}{f\left(|\vec{w}|\right)} \\
    &= \sup_{\substack{\vec{w} \in \mathbb{E}^{xz}_{p_{1}},\ |\vec{w}|<1}}
    (1-q)
    \frac{\arctanh\left(|\vec{w}_{\mathcal{A}_{q}}|\right)}{|\vec{w}_{\mathcal{A}_{q}}|}
    \frac{|\vec{w}|}{\arctanh\left(|\vec{w}|\right)} .
\end{align}

Let $\vec{w}=(w_{1},0,w_{3})=(r \cos\theta, 0, r \sin\theta)$ and hence $\vec{w}_{\mathcal{A}_{q}}=(\sqrt{1-q}\,w_{1},0,q+(1-q)w_{3})=(\sqrt{1-q}\, r \cos\theta, 0,q+(1-q)r \sin\theta)$.
 
Define $m_{r}(\theta)=|\vec{w}_{\mathcal{A}_{q}}|$.
Then 
\begin{align}
    \Pi(q)
    &= \sup_{\substack{r,\theta}}
    (1-q)
    \frac{\arctanh\left(m_{r}(\theta)\right)}{m_{r}(\theta)}
    \frac{r}{\arctanh(r)} \\
    &= \sup_{\substack{r}}
    (1-q)
    \frac{r}{\arctanh(r)}
    \sup_{\substack{\theta}}
    \frac{\arctanh\left(m_{r}(\theta)\right)}{m_{r}(\theta)} ,
\end{align}
where $(r,\theta)$ ranges over the pairs with $(r\cos\theta,0,r\sin\theta)\in\mathbb{E}^{xz}_{p_{1}}$ and $r<1$.

It can be calculated that 
\begin{align}
    m_{r}^{2}(\theta)=q^{2}+2q(1-q)r\sin\theta+(1-q)r^{2}\left(1-q\sin^{2}\theta\right) .
\end{align}
Viewing $m_{r}^{2}(\theta)$ as a function of $\sin\theta$, exactly as in the proof of Lemma~\ref{lem:yaxis}, we have $m_{r}(\theta)\leq m_{r}(\frac{\pi}{2})=q+(1-q)r$.

It remains to verify the admissible radii and
the feasibility of $\theta=\pi/2$.
Since
\begin{align}
    \mathbb{E}_{p_1}^{xz}\cap\{w_1=0\}
    =\{(0,0,z):2p_1-1\leq z\leq1\},
\end{align}
we have $(0,0,r)\in\mathbb{E}_{p_1}^{xz}$
for every $r\in I_{p_1}$.
Conversely, if $p_1>1/2$, every
$\vec{w}\in\mathbb{E}_{p_1}^{xz}$ satisfies
\begin{align}
    |\vec{w}|\geq w_3\geq2p_1-1.
\end{align}
Thus the admissible radii with $|\vec{w}|<1$
are precisely $I_{p_1}$, and
$\theta=\pi/2$ is feasible for every such
radius.

Combining these facts with the fact that $\frac{\arctanh(x)}{x}$ is increasing in $x\in[0,1)$, we obtain the following result:
\begin{align}
    \Pi(q)=
    \sup_{\substack{r\in I_{p_{1}}}}
    (1-q)
    \frac{r}{\arctanh(r)}
    \frac{\arctanh\left(q+(1-q)r\right)}{q+(1-q)r}.
\end{align}
%where the supremum is taken over $[0,1)$ for $p_{1} \in (0,\frac{1}{2}]$ and over $[2p_{1}-1,1)$ for $p_{1} \in (\frac{1}{2},1)$.
    
\subsection{Proof of Theorem~\ref{thm:relexpansion}}
\label{sec:relexpproof}

By Lemma~\ref{lemma:BKM}, it suffices to prove
\begin{align}
    \check{\eta}_{\mathrm{BKM}}
    \left(\mathcal{A}_{p_{2}},\mathcal{A}_{p_{1}}\right)
    &:=
    \inf_{\substack{\rho \in \mathcal{D}(\mathcal{X}) \\
    X=X^{\dagger},\ \mathrm{Tr}(X)=0 \\
    X\neq 0}}
    \frac
    {g_{\mathcal{A}_{p_{2}}(\rho)}\left(\mathcal{A}_{p_{2}}(X)\right)}
    {g_{\mathcal{A}_{p_{1}}(\rho)}\left(\mathcal{A}_{p_{1}}(X)\right)} \\
    &= \frac{p_{1}(1-p_{2})}{p_{2}(1-p_{1})}.
\end{align}
As in the proof of Theorem~\ref{thm:relcontraction}, the above optimization is equivalent to
\begin{align}
    \check{\eta}_{\mathrm{BKM}}
    \left(\mathcal{A}_{p_{2}},\mathcal{A}_{p_{1}}\right)
    =
    \inf_{\substack{
    \omega\in\mathcal{A}_{p_{1}}\left(\mathcal{D}(\mathcal{X})\right) \\
    Y=Y^{\dagger},\ \mathrm{Tr}(Y)=0 \\
    Y\neq 0}}
    \frac
    {g_{\mathcal{A}_{q}(\omega)}\left(\mathcal{A}_{q}(Y)\right)}
    {g_{\omega}(Y)} ,
\end{align}
and it suffices to restrict the Bloch vector $\vec{w}$ to $\Pxz$.
Therefore, we only need to consider $\vec{w}=(w_{1},0,w_{3})\in \mathbb{E}^{xz}_{p_{1}}$.
Consequently, the relative expansion problem is reduced to computing
\begin{align}
    \Omega(q) =
    \inf_{\substack{ \vec{w}\in \mathbb{E}^{xz}_{p_{1}},\ |\vec{w}|<1 \\ \vec{v}\neq \vec{0}}}
    \frac{g_{\vec{w}_{\mathcal{A}_{q}}}\left(\vec{v}_{\mathcal{A}_{q}}\right)}
    {g_{\vec{w}}(\vec{v})} .
\end{align}

We also decompose $g_{\vec{w}}(\vec{v})$ and $g_{\vec{w}_{\mathcal{A}_{q}}}(\vec{v}_{\mathcal{A}_{q}})$ into two parts, respectively. When $\vec{v}$ lies in $\Py$, we obtain
\begin{align}
    \inf_{\substack{\vec{w} \in \mathbb{E}^{xz}_{p_{1}},\ |\vec{w}|<1 \\ \vec{v} \in \Py,\ \vec{v}\neq\vec{0}}}
    \frac{g_{\vec{w}_{\mathcal{A}_{q}}}(\vec{v}_{\mathcal{A}_{q}})}{g_{\vec{w}}(\vec{v})}
    &= \inf_{\substack{\vec{w} \in \mathbb{E}^{xz}_{p_{1}},\ |\vec{w}|<1}}
    (1-q)
    \frac{1-|\vec{w}|^{2}}{1-|\vec{w}_{\mathcal{A}_{q}}|^{2}}
    \frac{f\left(|\vec{w}_{\mathcal{A}_{q}}|\right)}{f\left(|\vec{w}|\right)} .
\end{align}
By the proof of Theorem~\ref{thm:relcontraction}, the eigenvalues corresponding to the restriction of $M^{-1}N$ to $\Pxz$ are $\lambda_{+}\left(M^{-1}N\right)$ and $\lambda_{-}\left(M^{-1}N\right)$. In particular, the eigenvector associated with $\lambda_{-}\left(M^{-1}N\right)$ lies in $\Pxz$.
Thus, we have
\begin{align}
    \inf_{\substack{\vec{w} \in \mathbb{E}^{xz}_{p_{1}},\ |\vec{w}|<1\\ \vec{v} \in \Pxz,\ \vec{v}\neq\vec{0}}}\frac{g_{\vec{w}_{\mathcal{A}_{q}}}(\vec{v}_{\mathcal{A}_{q}})}{g_{\vec{w}}(\vec{v})}
    =\inf_{\substack{\vec{w} \in \mathbb{E}^{xz}_{p_{1}},\ |\vec{w}|<1}} \lambda_{-}(M^{-1}N).
\end{align}

Since $\lambda_{-}\leq\lambda_{+}\leq\lambda_{y}$ by~\eqref{eq:cd} (with $p$ replaced by $q$), and since $\frac{a_{1}+a_{2}}{b_{1}+b_{2}}\geq\min\{\frac{a_{1}}{b_{1}},\frac{a_{2}}{b_{2}}\}$ for $a_{i}\geq0$, $b_{i}>0$, the decomposition $g=g(\cdot_{y})+g(\cdot_{xz})$ shows that, for each fixed $\vec{w}$, the infimum over all $\vec{v}\neq\vec{0}$ equals $\lambda_{-}(M^{-1}N)$.
In conclusion, 
\begin{align}
    \Omega(q)
    &= \inf_{\substack{\vec{w} \in \mathbb{E}^{xz}_{p_{1}},\ |\vec{w}|<1}} \lambda_{-}(M^{-1}N) \\
    &= \inf_{\substack{\vec{w} \in \mathbb{E}^{xz}_{p_{1}},\ |\vec{w}|<1}}
    \frac{1-|\vec{w}|^{2}}{1-|\vec{w}_{\mathcal{A}_{q}}|^{2}}
    \frac{\Lambda_{\vec{w}}-\sqrt{\Lambda_{\vec{w}}^{2}-4(1-q)^{3}f\left(|\vec{w}|\right)f\left(|\vec{w}_{\mathcal{A}_{q}}|\right)}}
    {2f\left(|\vec{w}|\right)}
\end{align}
where 
\begin{align}
    \Lambda_{\vec{w}}
    &=
    \frac{f(|\vec{w}|)w_{3}^{2}+w_{1}^{2}}{|\vec{w}|^{2}}
    (1-q)^{2}
    \frac
    {f(|\vec{w}_{\mathcal{A}_{q}}|)\widetilde{w}_{1}^{2}+\widetilde{w}_{3}^{2}}
    {|\vec{w}_{\mathcal{A}_{q}}|^{2}} \\
    &\quad
    +
    \frac{f(|\vec{w}|)w_{1}^{2}+w_{3}^{2}}{|\vec{w}|^{2}}
    (1-q)
    \frac{f(|\vec{w}_{\mathcal{A}_{q}}|)\widetilde{w}_{3}^{2}+\widetilde{w}_{1}^{2}}
    {|\vec{w}_{\mathcal{A}_{q}}|^{2}} \\
    &\quad
    -
    2
    \frac{\left(1-f(|\vec{w}|)\right)w_{1}w_{3}}{|\vec{w}|^{2}}
    (1-q)^{\frac{3}{2}}
    \frac
    {\left(1-f(|\vec{w}_{\mathcal{A}_{q}}|)\right)\widetilde{w}_{1}\widetilde{w}_{3}}
    {|\vec{w}_{\mathcal{A}_{q}}|^{2}} ,
    \end{align}
with $\widetilde{w}_{1}=\sqrt{1-q}\,w_{1}$ and $\widetilde{w}_{3}=q+(1-q)w_{3}$.
    
It can be calculated that 
\begin{align}
    \inf_{\substack{\vec{w} \in \mathbb{E}^{xz}_{p_{1}},\ |\vec{w}|<1}}
    \lambda_{-}(M^{-1}N)
    &= \frac{p_{1}(1-p_{2})}{p_{2}(1-p_{1})} , \label{eq:cg}
\end{align}
whose proof is deferred to Appendix~\ref{ap:E}.

\appendix

\section{Proof of~\texorpdfstring{\eqref{eq:ca}}{Equation \ref*{eq:ca}}}
\label{ap:A}

In this appendix, we prove that
\begin{align}
    \eta(\mathcal{D}) \leq 4p(1-p)\frac{\arctanh\left(|1-2p|\right)}{|1-2p|}=f(|1-2p|).
\end{align}
Since $0<f(z)<1$ for $z\in(0,1)$ (see~\eqref{eq:fseries}), equivalently $(1-z^{2})\arctanh(z)<z$, it follows that $\eta(\mathcal{D})<1$ for $p\in(0,\frac{1}{2})\cup(\frac{1}{2},1)$. 

Let $\mathcal{X}=\mathbb{C}^{\{0,1\}}$.
The completely dephasing channel is defined by
\begin{align}
    \Delta(X)=\sum_{i=0}^{1}\bra{i}X\ket{i}\ket{i}\bra{i},
    \qquad X\in\mathcal{L}(\mathcal{X}).
\end{align}
Since $\mathcal{D}=\mathcal{D}\circ\Delta$,
pairs satisfying $\Delta(\rho)=\Delta(\sigma)$
give a zero numerator and can be discarded.
Thus, by the data-processing inequality,
\begin{align}
\eta(\mathcal{D})
&=\sup_{\substack{\rho\ll\sigma\\
\Delta(\rho)\neq\Delta(\sigma)}}
\frac{D(\mathcal{D}(\rho)\|
\mathcal{D}(\sigma))}
{D(\rho\|\sigma)}\\
&=\sup_{\substack{\rho\ll\sigma\\
\Delta(\rho)\neq\Delta(\sigma)}}
\frac{D(\mathcal{D}(\Delta(\rho))\|
\mathcal{D}(\Delta(\sigma)))}
{D(\rho\|\sigma)}\\
&\leq\sup_{\substack{\rho\ll\sigma\\
\Delta(\rho)\neq\Delta(\sigma)}}
\frac{D(\mathcal{D}(\Delta(\rho))\|
\mathcal{D}(\Delta(\sigma)))}
{D(\Delta(\rho)\|\Delta(\sigma))}\\
&\leq\sup_{\substack{\rho\neq\sigma,\ \rho\ll\sigma\\
\rho,\sigma\ \mathrm{diagonal}}}
\frac{D(\mathcal{D}(\rho)\|
\mathcal{D}(\sigma))}
{D(\rho\|\sigma)}.
\end{align}

Let
\begin{align}
    \rho=
    \begin{pmatrix}
        a & 0 \\
        0 & 1-a
    \end{pmatrix},
    \quad
    \sigma=
    \begin{pmatrix}
        b & 0 \\
        0 & 1-b
    \end{pmatrix},
    \quad 
    a,b \in [0,1].
\end{align}
For such diagonal states, the condition $\rho \ll \sigma$ is equivalent to $\mathrm{supp}(\rho) \subseteq \mathrm{supp}(\sigma)$. In particular, if $a\in(0,1)$, then $\rho$ has full support. Hence $\rho \ll \sigma$ holds if and only if $\sigma$ also has full support, that is, $b\in(0,1)$.
Define two states $\rho_{0}=\begin{pmatrix} 1-p & -\sqrt{p(1-p)} \\ -\sqrt{p(1-p)} & p \end{pmatrix}$ and $\rho_{1}=\begin{pmatrix} 1-p & \sqrt{p(1-p)} \\ \sqrt{p(1-p)} & p \end{pmatrix}$. Then $\mathcal{D}(\rho)=(1-a)\rho_{0}+a\rho_{1}$. For simplicity, write
\begin{align}
    M(a):=\mathcal{D}(\rho)=
    \begin{pmatrix}
    1-p & (2a-1)\sqrt{p(1-p)} \\
    (2a-1)\sqrt{p(1-p)} & p
    \end{pmatrix},
\end{align}
and similarly write $M(b):=\mathcal{D}(\sigma)$.

Use the notation $D(a\|b):=a\log\frac{a}{b}+(1-a)\log\frac{1-a}{1-b}$. Then the ratio $\frac{D\left(M(a)\|M(b)\right)}{D(a\|b)}$ is continuous on the set $\Omega:=\{(a,b)\in(0,1)^{2}: a \neq b\}$. 
Therefore, it remains to prove that 
\begin{align}
    \sup\limits_{\substack{ (a,b) \in \Omega}}\frac{D\left(M(a)\|M(b)\right)}{D(a\|b)}<1.
\end{align}

Motivated by the Bregman-divergence representation of relative entropy~\cite{ref42}, define
\begin{align}
    F(a)=\mathrm{Tr}\left[M(a)\log M(a)\right], \quad 
    G(a)=a\log a+(1-a)\log(1-a).
\end{align}
Then, for $a,b\in(0,1)$, we have
\begin{align}
    D(a\|b)=G(a)-G(b)-G'(b)(a-b),
\end{align}
and, applying the same Bregman-divergence representation to the affine family $M(a)$,
\begin{align}
    D\left(M(a) \| M(b)\right)=F(a)-F(b)-F'(b)(a-b).
\end{align} 

For any $C^{2}$ function $H$, one has 
\begin{align}
    H(a)-H(b)-H'(b)(a-b)
    &=\int_{b}^{a}\left[H'(x)-H'(b)\right]dx \\
    &=\int_{b}^{a}\int_{b}^{x}H''(y)\,dy\,dx .
\end{align}
Hence, we can infer that $D(M(a)\|M(b)) \leq C \cdot D(a\|b)$ from $F''(t) \leq C \cdot G''(t)$ for $t\in(0,1)$, where $C$ is a constant.
Hence, the desired claim reduces to proving $\sup\limits_{t\in(0,1)}\frac{F''(t)}{G''(t)}<1$.

Let $z:=|1-2p|$ and, for $t\in(0,1)$, let $s:=2t-1$ and $u:=\sqrt{z^{2}+(1-z^{2})s^{2}}\in[z,1)$. A direct calculation shows that the eigenvalues of $M(t)$ are $\lambda_{1,2}(t)=\frac{1 \pm u}{2}$.
Hence, 
\begin{align}
    F(t)
    &=\frac{1+u}{2}\log\frac{1+u}{2}+\frac{1-u}{2}\log\frac{1-u}{2} \\
    &=\frac{1}{2}\log\frac{1-u^{2}}{4}+u\cdot \arctanh(u).
\end{align}
A straightforward calculation yields
\begin{align}
    \frac{F''(t)}{G''(t)}=
    (1-z^{2})\left[\frac{z^{2}(1-s^{2})}{u^{3}}\arctanh(u)+\frac{s^{2}}{u^{2}}\right] .
\end{align}

Using $(1-z^{2})s^{2}=u^{2}-z^{2}$ and $(1-z^{2})(1-s^{2})=1-u^{2}$, we can write 
\begin{align}
    \Gamma(u)
    &:= \frac{F''(t)}{G''(t)} \\
    &= 1-\frac{z^{2}}{u^{2}}+\frac{z^{2}(1-u^{2})}{u^{3}}\arctanh(u)
\end{align}
as a function of $u \in [z,1)$. Next we prove that $\Gamma$ is decreasing on $(0,1)$ and hence 
\begin{align}
    \Gamma(u)
    \leq \Gamma(z)
    = 4p(1-p)\frac{\arctanh\left(|1-2p|\right)}{|1-2p|} .
\end{align}

A direct differentiation gives $\Gamma'(u)=\frac{z^{2}}{u^{4}}\left(u^{2}\,\arctanh(u)+3u-3\arctanh(u)\right)$. 
Thus, it remains to determine the sign of $h(u):=u^{2}\,\arctanh(u)+3u-3\arctanh(u)$.
Differentiating $h$, we obtain
\begin{align}
    h'(u)
    &=2u \, \arctanh(u)+\frac{u^{2}}{1-u^{2}}+3-\frac{3}{1-u^{2}} \\
    &=2u \, \arctanh(u)-\frac{2u^{2}}{1-u^{2}} \\
    &=2u\left(\arctanh(u)-\frac{u}{1-u^{2}}\right).
\end{align}
For $0<u<1$, we have
\begin{align}
    \arctanh(u)
    =\int_{0}^{u}\frac{dt}{1-t^{2}}
    <\int_{0}^{u}\frac{dt}{1-u^{2}}
    =\frac{u}{1-u^{2}}.
\end{align}

It follows that $h'(u)<0$ for all $u\in(0,1)$. 
Since $h(0)=0$, we conclude that $h(u)<0$ and hence $\Gamma'(u)<0$ for all $u\in(0,1)$. 
Consequently, 
\begin{align}
    \sup_{t\in(0,1)}\frac{F''(t)}{G''(t)}=\sup_{u\in[z,1)}\Gamma(u)=\Gamma(z)=f(z)<1,
\end{align}
which proves~\eqref{eq:ca}.

\section{Proof of~\texorpdfstring{\eqref{eq:cb}}{Equation \ref*{eq:cb}}}
\label{ap:B}

For $r\in[0,1)$, let $s=p+(1-p)r$. Then $r<s<1$ for $r\in[0,1)$, and $1-s=(1-p)(1-r)$.
Define
\begin{align}
    \varphi(x)
    =
    \frac{1-x}{x}\arctanh(x),
    \quad x\in(0,1)
\end{align}
with the convention $\varphi(0)=1$. 
Hence we have
\begin{align}
    \vartheta_{p}(r)
    &=
    (1-p)
    \frac{r}{\arctanh(r)}
    \frac{\arctanh(s)}{s} \\
    &=
    \frac{1-s}{1-r}
    \frac{r}{\arctanh(r)}
    \frac{\arctanh(s)}{s} \\
    &=
    \frac{\varphi(s)}{\varphi(r)} ,
\end{align}
for $r\in(0,1)$.

We claim that $\varphi$ is strictly decreasing on $[0,1)$. Indeed, a direct computation gives 
\begin{align}
    \varphi'(x)
    &= \frac{1}{x(1+x)}-\frac{\arctanh(x)}{x^{2}} .
\end{align}
Since
\begin{align}
    \arctanh(x)
    =\int_{0}^{x}\frac{1}{1-u^{2}}du
    >\int_{0}^{x}\frac{1}{1+x}du
    =\frac{x}{1+x},
\end{align}
we have $\varphi'(x)<0$ for all $x\in(0,1)$. Hence $\varphi$ is strictly decreasing. Since $s>r$, it follows that 
\begin{align}
    \vartheta_{p}(r)=\frac{\varphi(s)}{\varphi(r)}<1
\end{align}
for every $r\in(0,1)$.

Moreover, $\vartheta_{p}$ has finite limits at the endpoints. 
At $r=0$,
\begin{align}
\vartheta_{p}(0)
=(1-p)\frac{\arctanh(p)}{p}
=\frac{\varphi(p)}{\varphi(0)}
<1 .
\end{align}
At the other endpoint,
\begin{align}
\lim_{r \to 1^{-}}\vartheta_{p}(r)
&=\lim_{r\to 1^{-}}
(1-p)
\frac{r}{p+(1-p)r}
\frac{\arctanh\left(p+(1-p)r\right)}{\arctanh(r)} \\
&=1-p \\
&<1 .
\end{align}
Therefore $\vartheta_{p}$ extends continuously to $[0,1]$ and is strictly less than $1$ on this compact interval.
 Hence we derive that 
\begin{align}
\sup_{r\in[0,1)}\vartheta_{p}(r)<1 .
\end{align}

\section{Proof of~\texorpdfstring{\eqref{eq:cc}}{Equation \ref*{eq:cc}}}
\label{ap:C}

In this appendix, we prove that for any $p\in(0,1)$ and any
$\vec{w}\in\mathbb{R}^{3}$ with $|\vec{w}|<1$, the transformed Bloch vector
$\vec{w}_{\mathcal{A}_{p}}$ also satisfies $|\vec{w}_{\mathcal{A}_{p}}|<1$.

Recall that the qubit AD channel $\mathcal{A}_{p}$ acts on the
Bloch vector $\vec{w}=(w_{1},w_{2},w_{3})$ as $\vec{w}_{\mathcal{A}_{p}}=\left(\sqrt{1-p}\,w_{1},\sqrt{1-p}\,w_{2},p+(1-p)w_{3}\right)$. Therefore, 
\begin{align*}
|\vec{w}_{\mathcal{A}_{p}}|^{2}
&=(1-p)w_{1}^{2}+(1-p)w_{2}^{2}+\left(p+(1-p)w_{3}\right)^{2} \\
&=(1-p)(w_{1}^{2}+w_{2}^{2})+p^{2}+2p(1-p)w_{3}+(1-p)^{2}w_{3}^{2}.
\end{align*}
Hence,
\begin{align*}
1-|\vec{w}_{\mathcal{A}_{p}}|^{2}
&=1-(1-p)(w_{1}^{2}+w_{2}^{2})-p^{2}-2p(1-p)w_{3}-(1-p)^{2}w_{3}^{2} \\
&=(1-p)\left[1+p-(w_{1}^{2}+w_{2}^{2})-2pw_{3}-(1-p)w_{3}^{2}\right]  \\
&=(1-p)\left[1-|\vec{w}|^{2}+p(1-w_{3})^{2}\right] .
\end{align*}
Since $p\in(0,1)$, we have $1-p>0$. 
Moreover, $|\vec{w}|<1$ implies $1-|\vec{w}|^{2}>0$,
and clearly $p(1-w_{3})^{2}\geq 0$.
It follows that
\begin{align*}
1-|\vec{w}_{\mathcal{A}_{p}}|^{2}
=(1-p)\left[1-|\vec{w}|^{2}+p(1-w_{3})^{2}\right]
>0 .
\end{align*}
Consequently, we derive that $|\vec{w}_{\mathcal{A}_{p}}|<1$.

\section{Proof of~\texorpdfstring{\eqref{eq:cd}}{Equation \ref*{eq:cd}}}\label{ap:D}

In this appendix we prove the spectral comparison used in the proof of the contraction coefficient for the AD channel. Throughout this appendix, let $\bar{p}=1-p$.

Recall that, for $\vec{w}=\left(w_{1},0,w_{3}\right)$, we have $\vec{w}_{\mathcal{A}_{p}}=\left(\widetilde{w}_{1},0,\widetilde{w}_{3}\right)$, where $\widetilde{w}_{1}=\sqrt{\bar{p}}\,w_{1}$ and $\widetilde{w}_{3}=p+\bar{p} w_{3}$.
Let $r=|\vec{w}|$ and $\tilde{r}=|\vec{w}_{\mathcal{A}_{p}}|$.
All formulas below are understood by continuity at $r=0$ or $\tilde{r}=0$.
For simplicity, write $f_{r}=f(r)$ and $\tilde{f}=f\left(\tilde{r}\right)$. 
Hence, from the block structure of $M$ and $N$, the eigenvalue corresponding to the $y$-direction is 
\begin{align}
    \lambda_{y}
    =
    \bar{p}\,
    \frac{1-r^{2}}{1-\tilde{r}^{2}}
    \frac{\tilde{f}}{f_{r}}.
    \label{eq:appD-lambday}
\end{align}
It remains to show that the two generalized eigenvalues coming from $\Pxz$ are bounded above by $\lambda_{y}$.

Let
\begin{align}
    B
    =
    \begin{pmatrix}
    \dfrac{f_{r}w_{3}^{2}+w_{1}^{2}}{r^{2}}
    &
    \dfrac{\left(1-f_{r}\right)w_{1}w_{3}}{r^{2}}
    \\
    \dfrac{\left(1-f_{r}\right)w_{1}w_{3}}{r^{2}}
    &
    \dfrac{f_{r}w_{1}^{2}+w_{3}^{2}}{r^{2}}
    \end{pmatrix}
\end{align}
and
\begin{align}
    C
    =
    \begin{pmatrix}
    \bar{p}
    \dfrac{\tilde{f}\widetilde{w}_{3}^{2}+\widetilde{w}_{1}^{2}}{\tilde{r}^{2}}
    &
    \bar{p}^{\frac{3}{2}}
    \dfrac{\left(1-\tilde{f}\right)\widetilde{w}_{1}\widetilde{w}_{3}}{\tilde{r}^{2}}
    \\
    \bar{p}^{\frac{3}{2}}
    \dfrac{\left(1-\tilde{f}\right)\widetilde{w}_{1}\widetilde{w}_{3}}{\tilde{r}^{2}}
    &
    \bar{p}^{2}
    \dfrac{\tilde{f}\widetilde{w}_{1}^{2}+\widetilde{w}_{3}^{2}}{\tilde{r}^{2}}
    \end{pmatrix}.
\end{align}
Then the $xz$-blocks of $M$ and $N$ are respectively $\frac{1}{1-r^{2}}B$ and $\frac{1}{1-\tilde{r}^{2}}C$. Therefore, the two eigenvalues of $M^{-1}N$
corresponding to $\Pxz$ are
\begin{align}
    \lambda_{\pm}
    =
    \frac{1-r^{2}}{1-\tilde{r}^{2}}
    \mu_{\pm},
\end{align}
where $\mu_{\pm}$ are the two eigenvalues of $A=B^{-1}C$.
Thus it suffices to prove
\begin{align}
    \mu_{\pm}
    \leq
    \mu_{y},
    \quad
    \mu_{y}
    =
    \bar{p}\,
    \frac{\tilde{f}}{f_{r}}.
\label{eq:appD-muy}
\end{align}

A direct computation gives $\det B=f_{r}$ and $\det C=\bar{p}^{3}\tilde{f}$. 
Hence $\det A=\bar{p}^{3}\frac{\tilde{f}}{f_{r}}=\mu_{y}\mu_{r}$, where $\mu_{r}=\bar{p}^{2}$. 
We next show that $\mu_{r}\leq\mu_{y}$, which will be used later.

\noindent{\bf Proof of $\mu_{r}\leq\mu_{y}$:}

For this purpose, we need a simple estimate on $\tilde{r}$. 
Since $\tilde{r}^{2}=\bar{p} w_{1}^{2}+\left(p+\bar{p} w_{3}\right)^{2}$, a direct computation gives
\begin{align}
    \left(p+\bar{p} r\right)^{2}-\tilde{r}^{2}
    =
    p\,\bar{p}\left(r-w_{3}\right)\left(2-r-w_{3}\right)
    \geq
    0.
\end{align}
Thus $\tilde{r} \leq p+\bar{p} r$.

Moreover, it follows from 
\begin{align}
    \arctanh x
    =\int_{0}^{x}\frac{dt}{1-t^{2}}
    \leq\int_{0}^{x}\frac{dt}{1-x}
    =\frac{x}{1-x}
\end{align}
that 
\begin{align}
    \frac{d}{dx}\log f(x)+\frac{1}{1-x}
    &=\frac{1}{\left(1-x^{2}\right)\arctanh x}-\frac{1}{x}+\frac{1}{1+x} \\
    &=\frac{1}{\left(1-x^{2}\right)\arctanh x}-\frac{1}{x\left(1+x\right)} \\
    &\geq 0 .
\end{align}
Integrating from $r$ to $p+\bar{p}r$ gives
\begin{align}
    f\left(p+\bar{p} r\right)
    \geq
    \bar{p} f(r).
\end{align}
Since $f$ is decreasing (see~\eqref{eq:fseries}) and $\tilde{r}\leq p+\bar{p} r$, we obtain
\begin{align}
    \tilde{f}
    =
    f\left(\tilde{r}\right)
    \geq
    f\left(p+\bar{p} r\right)
    \geq
    \bar{p} f_{r}.
\end{align}
Consequently,
\begin{align}
    \mu_{y}=\bar{p}\frac{\tilde{f}}{f_{r}}\geq\bar{p}^{2}=\mu_{r}.
\end{align}

It remains to prove $\mu_{+}\leq \mu_{y}$. Define
\begin{align}
    n(x)=\frac{1-f(x)}{x^{2}}, \quad 0<x<1,
\end{align}
with the continuous extension $n(0)=\frac{2}{3}$. Then we have
\begin{align}
    n(x)
    =\frac{1-\left(1-x^{2}\right)\dfrac{\arctanh x}{x}}{x^{2}}
    =\int_{0}^{1}\frac{1-t^{2}}{1-x^{2}t^{2}}dt.
\label{eq:appD-n-integral}
\end{align}
In particular, $0<n(x)<1$ and $n$ is increasing on $(0,1)$.

The following scalar inequality is the key point:
\begin{align}
    n\left(\tilde{r}\right)
    \leq
    \frac{n(r)}{\bar{p}+p\,n(r)}. \label{eq:appD-scalar}
\end{align}
To prove it, define
\begin{align}
    \ell(x)
    =\frac{1}{n(x)}-1
    =\frac{1-n(x)}{n(x)}.
\end{align}
Since $n$ is increasing and $0<n<1$, the function $\ell$ is positive and decreasing on $(0,1)$.

\noindent{\bf Proof of~\eqref{eq:appD-scalar}:}

We claim that
\begin{align}
    \frac{d}{dx}\log \ell(x)
    \geq -\frac{1}{1-x}.
\label{eq:appD-logell}
\end{align}
Then integrating~\eqref{eq:appD-logell} from $r$ to $p+\bar{p} r$, we get
\begin{align}
    \ell\left(p+\bar{p} r\right)
    \geq
    \bar{p} \ell(r).
\end{align}
Since $\ell$ is decreasing and $\tilde{r}\leq p+\bar{p} r$, it follows that
\begin{align}
    \ell\left(\tilde{r}\right)
    \geq \ell\left(p+\bar{p} r\right)
    \geq \bar{p}\ell(r),
\end{align}
which is equivalent to
\begin{align}
    n\left(\tilde{r}\right)
    \leq
    \frac{n(r)}{\bar{p}+p\, n(r)}.
\end{align}

Indeed, from~\eqref{eq:appD-n-integral}, let
\begin{align}
    \phi_{x}(t)
    =
    \frac{1-t^{2}}{1-x^{2}t^{2}},
    \quad
    \psi_{x}(t)
    =
    \frac{\left(1-x^{2}\right)t^{2}}{1-x^{2}t^{2}} .
\end{align}
Then we obtain $\phi_{x}(t)+\psi_{x}(t)=1$.
Moreover,
\begin{align}
    n'(x)
    =\int_{0}^{1}\frac{2xt^{2}\left(1-t^{2}\right)}{\left(1-x^{2}t^{2}\right)^{2}}dt
    =\frac{2x}{1-x^{2}}\int_{0}^{1}\phi_{x}(t)\psi_{x}(t)dt.
\end{align}
The function $\phi_{x}(t)$ is decreasing in $t$, while $\psi_{x}(t)$ is increasing in $t$. Hence, by the reverse Chebyshev inequality,
\begin{align}
    \int_{0}^{1}\phi_{x}(t)\psi_{x}(t)dt
    \leq 
    \left(\int_{0}^{1}\phi_{x}(t)dt\right)\left(\int_{0}^{1}\psi_{x}(t)dt\right)
    =
    n(x)\left(1-n(x)\right).
\end{align}
Since $\frac{2x}{1+x}\leq 1$ for $0\leq x<1$,
\begin{align}
    n'(x)
    \leq
    \frac{2x}{1-x^{2}}n(x)\left(1-n(x)\right)
    \leq
    \frac{1}{1-x}n(x)\left(1-n(x)\right).
\end{align}
Therefore,
\begin{align}
    \frac{d}{dx}\log \ell(x)
    = -\frac{n'(x)}{n(x)\left(1-n(x)\right)}
    \geq -\frac{1}{1-x},
\end{align}
which proves~\eqref{eq:appD-logell}.

We now finish the comparison. The characteristic polynomial of $A$ is
\begin{align}
    P_{A}(\mu)=\mu^{2}-\left(\mathrm{tr}A\right)\mu+\det A.
\end{align}
We will prove $P_A(\mu_y)\geq 0$ below.
Since
\[
P_A(\mu)=(\mu-\mu_-)(\mu-\mu_+),
\qquad 0<\mu_-\leq\mu_+,
\]
we have $\mu_y\leq\mu_-$ or $\mu_y\geq\mu_+$.
In the first case, using $\mu_r\leq\mu_y$
and $\det A=\mu_y\mu_r=\mu_-\mu_+$,
we obtain
\[
\mu_+
=\frac{\mu_y\mu_r}{\mu_-}
\leq\frac{\mu_y^2}{\mu_-}
\leq\mu_y
\leq\mu_-
\leq\mu_+.
\]
Thus equality holds throughout.
Consequently, $\mu_+\leq\mu_y$ in both cases.

\noindent{\bf Proof of $P_{A}(\mu_{y})\geq 0$:}

Notice that 
\begin{align}
    P_{A}\left(\mu_{y}\right)
    =\mu_{y}^{2}-\left(\mathrm{tr}A\right)\mu_{y}+\mu_{y}\mu_{r}
    =\mu_{y}\left(\mu_{y}+\mu_{r}-\mathrm{tr}A\right).
\end{align}
It suffices to prove $\mu_{y}+\mu_{r}-\mathrm{tr}A\geq0$. 
For simplicity, write $n_{r}=n(r)$ and $\tilde{n}=n\left(\tilde{r}\right)$.
A direct calculation gives 
\begin{align}
    1-f_{r}=r^{2}n_{r}, \quad 1-\tilde{f}=\tilde{r}^{2}\tilde{n} . \label{eq:de}
\end{align}
Combining~\eqref{eq:de} with the equality $1-\tilde{r}^{2}=\bar{p}\left[1-r^{2}+p\left(1-w_{3}\right)^{2}\right]$ gives
\begin{align}
    \mu_{y}+\mu_{r}-\mathrm{tr}A
    &= \frac{p\bar{p} w_{1}^{2}}{f_{r}r^{2}\tilde{r}^{2}}
    \left[
    \bar{p} r^{2}\left(\tilde{f}-f_{r}\right)+p\left(1-f_{r}\right)\left(\tilde{f}-\bar{p}\left(1-w_{3}\right)^{2}\right)
    \right] \\
    &= \frac{p\bar{p} w_{1}^{2}}{f_{r}}
    \left[n_{r}-\left(\bar{p}+p\, n_{r}\right)\tilde{n}\right] \\
    &\geq 0 ,
\end{align}
where the last inequality comes from~\eqref{eq:appD-scalar}.

\section{Proof of~\texorpdfstring{\eqref{eq:cg}}{Equation \ref*{eq:cg}}}
\label{ap:E}

Throughout this appendix, $\bar{q}:=1-q$, $\vec{w}=(w_{1},0,w_{3})\in\mathbb{E}^{xz}_{p_{1}}$ with $|\vec{w}|<1$, and $M$, $N$ are the matrices~\eqref{eq:ce} and~\eqref{eq:cf} with $p$ replaced by $q$.
From the block structure of $M$ and $N$ and the spectral comparison in Appendix~\ref{ap:D}, we have
\begin{align}
    \lambda_{-}(M^{-1}N)
    &= \inf\limits_{\substack{\vec{v} \in \Pxz,\ \vec{v} \neq \vec{0}}}
    \frac{\vec{v}^{T}N\vec{v}}{\vec{v}^{T}M\vec{v}} \\
    &= \inf\limits_{u\in\mathbb{R}^{2}\setminus\{0\}}
    \frac{u^{T}N_{xz}u}{u^{T}M_{xz}u},
\end{align}
where 
\begin{align}
    M_{xz}
    &=
    \frac{1}{1-|\vec{w}|^{2}}
    \begin{pmatrix}
    \dfrac{f\left(|\vec{w}|\right)w_{3}^{2}+w_{1}^{2}}{|\vec{w}|^{2}}
    & \dfrac{\left(1-f\left(|\vec{w}|\right)\right)w_{1}w_{3}}{|\vec{w}|^{2}} \\
    \dfrac{\left(1-f\left(|\vec{w}|\right)\right)w_{1}w_{3}}{|\vec{w}|^{2}} & \dfrac{f\left(|\vec{w}|\right)w_{1}^{2}+w_{3}^{2}}{|\vec{w}|^{2}}
    \end{pmatrix}
\end{align}
and
\begin{align}
    N_{xz}
    &=\frac{1}{1-|\vec{w}_{\mathcal{A}_{q}}|^{2}}
    \begin{pmatrix}
    \bar{q} \,\dfrac{f\left(|\vec{w}_{\mathcal{A}_{q}}|\right)\widetilde{w}_{3}^{2}+\widetilde{w}_{1}^{2}}{|\vec{w}_{\mathcal{A}_{q}}|^{2}} & \bar{q}^{\frac{3}{2}}
    \dfrac{\left(1-f\left(|\vec{w}_{\mathcal{A}_{q}}|\right)\right)\widetilde{w}_{1}\widetilde{w}_{3}}{|\vec{w}_{\mathcal{A}_{q}}|^{2}} \\
    \bar{q}^{\frac{3}{2}}
    \dfrac{\left(1-f\left(|\vec{w}_{\mathcal{A}_{q}}|\right)\right)\widetilde{w}_{1}\widetilde{w}_{3}}{|\vec{w}_{\mathcal{A}_{q}}|^{2}} & \bar{q}^{2}
    \dfrac{f\left(|\vec{w}_{\mathcal{A}_{q}}|\right)\widetilde{w}_{1}^{2}+\widetilde{w}_{3}^{2}}{|\vec{w}_{\mathcal{A}_{q}}|^{2}}
    \end{pmatrix} ,
\end{align}
with $\widetilde{w}_{1}=\sqrt{\bar{q}}\,w_{1}$ and $\widetilde{w}_{3}=q+\bar{q}w_{3}$.

For $\vec{a}=(a_{1},0,a_{3})$, let $|\vec{a}|=\sqrt{a_{1}^{2}+a_{3}^{2}}<1$.
Define the two-dimensional matrix
\begin{align}
    \mathcal{M}(\vec{a})
    =
    \begin{pmatrix}
    \dfrac{f(|\vec{a}|)a_{3}^{2}+a_{1}^{2}}{|\vec{a}|^{2}}
    &
    \dfrac{\left(1-f(|\vec{a}|)\right)a_{1}a_{3}}{|\vec{a}|^{2}}
    \\[1.2em]
    \dfrac{\left(1-f(|\vec{a}|)\right)a_{1}a_{3}}{|\vec{a}|^{2}}
    &
    \dfrac{f(|\vec{a}|)a_{1}^{2}+a_{3}^{2}}{|\vec{a}|^{2}}
    \end{pmatrix},
\end{align}
where at $\vec{a}=\vec{0}$ the matrix is understood by continuous extension, namely $\mathcal{M}(\vec{0})=I_{2}.$
Let
\begin{align}
    D_{q}
    =
    \mathrm{diag}\left(\sqrt{1-q},1-q\right).
\end{align}
Then we have 
\begin{align}
    M_{xz}=\frac{1}{1-|\vec{w}|^{2}}\,\mathcal{M}(\vec{w}),
    \quad
    N_{xz}=\frac{1}{1-|\vec{w}_{\mathcal{A}_{q}}|^{2}}\,D_{q}\mathcal{M}(\vec{w}_{\mathcal{A}_{q}})D_{q}.
\end{align}
Therefore
\begin{align}
    \lambda_{-}(M^{-1}N)
    =
    \frac{1-|\vec{w}|^{2}}{1-|\vec{w}_{\mathcal{A}_{q}}|^{2}}
    \min_{\vec{u}\neq \vec{0}}
    \frac
    {\vec{u}^{T}D_{q}\mathcal{M}(\vec{w}_{\mathcal{A}_{q}})D_{q}\vec{u}}
    {\vec{u}^{T}\mathcal{M}(\vec{w})\vec{u}}.
\end{align}

Next, we give a brief outline of the proof of $\inf\limits_{\substack{\vec{w} \in \mathbb{E}^{xz}_{p_{1}}}} \lambda_{-}(M^{-1}N)=\frac{p_{1}(1-p_{2})}{p_{2}(1-p_{1})}$.
\begin{itemize}
\item[(1)] 
We first prove that
\begin{align}
    \frac{1-|\vec{w}|^{2}}{1-|\vec{w}_{\mathcal{A}_{q}}|^{2}}
    \geq
    \frac{p_{1}}{(1-q)\widehat{q}},
\end{align}
where $\widehat{q}=q+p_{1}-qp_{1}$$\,=p_{2}$.

\item[(2)]
Then we prove
\begin{align}
    D_{q}\mathcal{M}(\vec{w}_{\mathcal{A}_{q}})D_{q}
    \geq
    (1-q)^{2}\mathcal{M}(\vec{w}). \label{eq:ea}
\end{align}

\item[(3)] 
Finally, we verify that the infimum is attained at $\vec{w}_{*}=(0,0,2p_{1}-1)$, namely, $\lambda_{-}(M^{-1}N)=\frac{p_{1}(1-p_{2})}{p_{2}(1-p_{1})}$ when $\vec{w}=(0,0,2p_{1}-1)$.

\end{itemize}

\noindent{\bf Proof of (1):}

Parametrize $\mathbb{E}^{xz}_{p_{1}}$ by
\begin{align}
    \vec{w}=\left(\sqrt{1-p_{1}}\,x,\,0,\,p_{1}+(1-p_{1})z\right),
    \quad
    x^{2}+z^{2} \leq 1,\ (x,z)\neq(0,1).
\end{align}
Then we have
\begin{align}
    \vec{w}_{\mathcal{A}_q}
=\bigl(\sqrt{1-\widehat{q}}\,x,\,0,\,\widehat{q}+(1-\widehat{q})z\bigr),
\qquad \widehat{q}=q+p_1-qp_1=p_2.
\end{align}
Let $R=1-x^{2}-z^{2}$ and $d=1-z$$\,>0$.
By Appendix~\ref{ap:C}, a simple calculation gives
\begin{align}
    \frac{1-|\vec{w}|^{2}}{1-|\vec{w}_{\mathcal{A}_{q}}|^{2}}
    &=
    \frac{(1-p_{1})(R+p_{1}d^{2})}
    {(1-\widehat q)(R+\widehat q d^{2})}.
\end{align}
Since $1-\widehat{q}=\bar{q}(1-p_{1})$, $R\geq0$ and $p_{1}<\widehat{q}$, we get
\begin{align}
    \frac{1-|\vec{w}|^{2}}{1-|\vec{w}_{\mathcal{A}_{q}}|^{2}}
    &= \frac{1}{\bar{q}}\frac{R+p_{1}d^{2}}{R+\widehat{q} d^{2}} \\
    &\geq \frac{1}{\bar{q}}\frac{p_{1}}{\widehat{q}}, \label{eq:eb}
\end{align}
with equality if and only if $R=0$.

\noindent{\bf Proof of (2):}

Let $\iota=-\log \bar{q}$.
Define
\begin{align}
    \vec{W}_{\tau}
    =
    \left(
    e^{-\frac{\tau}{2}}w_{1},\,0,\,1-e^{-\tau}+e^{-\tau}w_{3}
    \right),
    \quad
    P_{\tau}=\mathrm{diag}\left(e^{\frac{\tau}{2}},1\right) ,
    \qquad \tau\geq0 .
\end{align}
Then $\vec{W}_{\iota}=\vec{w}_{\mathcal{A}_{q}}$ and $\vec{W}_{0}=\vec{w}$. 
Since $P_{\iota}=e^{\iota}D_{q}$ and $P_{0}=I_{2}$, \eqref{eq:ea} is equivalent to $P_{\iota}\mathcal{M}(\vec{W}_{\iota})P_{\iota} \geq P_{0}\mathcal{M}(\vec{W}_{0})P_{0}$.
Hence, it is enough to prove that
\begin{align}
    H(\tau)
    :=
    P_{\tau}\mathcal{M}(\vec{W}_{\tau})P_{\tau}
\end{align}
is increasing in $\tau$ (in the Loewner order).

Write $\vec{W}_{\tau}=(r\sin\theta,0,r\cos\theta)$ and $f_{r}=f(r)$.
In the orthonormal basis $e_{1}=(\sin\theta,\cos\theta)$ and $e_{2}=(\cos\theta,-\sin\theta)$, the matrix $\mathcal{M}(\vec{W}_{\tau})$ is diagonal:
\begin{align}
    \mathcal{M}(\vec{W}_{\tau})
    =
    \begin{pmatrix}
    1 & 0 \\
    0 & f_{r}
    \end{pmatrix}.
\end{align}
For simplicity, write $S=\sin\theta$ and $Z=\cos\theta$.
Then a direct computation gives $P_{\tau}^{-1}H'(\tau)P_{\tau}^{-1}=K$, where, in the basis $\{e_{1},e_{2}\}$,
\begin{align}
    K=
    \begin{pmatrix}
    S^{2} & S\left(Z-\dfrac{1-f_{r}}{r}\right) \\
    S\left(Z-\dfrac{1-f_{r}}{r}\right) & f'(r)\left(Z-\dfrac{r}{2}(1+Z^{2})\right)+f_{r}Z^{2}
    \end{pmatrix}.
\end{align}
The polar representation is undefined at
$\vec W_\tau=0$, which can occur for at most
one value of $\tau$. Since $H$ is continuous,
it suffices to establish $K\geq0$ whenever
$\vec W_\tau\neq0$.

Let $B_{r}(Z)=f'(r)\left(Z-\frac{r}{2}(1+Z^{2})\right)+f_{r}Z^{2}-\left(Z-\frac{1-f_{r}}{r}\right)^{2}$. Then $\det K=S^{2}B_{r}(Z)$.

If $S\neq0$, then $K\geq0$ follows from $K_{11}=S^{2}>0$ and $\det K\geq0$; if $S=0$, then $K=\mathrm{diag}(0,K_{22})$ with $K_{22}=B_{r}(Z)+\left(Z-\frac{1-f_{r}}{r}\right)^{2}$.
Hence it is enough to prove $B_{r}(Z)\geq 0$ for $-1\leq Z\leq 1$ and $r\in(0,1)$.

Since
\begin{align}
    \frac{f(r)}{1-r^{2}}=\frac{\arctanh r}{r}=\int_{0}^{1}\frac{ds}{1-r^{2}s^{2}},
\end{align} 
Jensen's inequality (applied to the convex function $u\mapsto(1-r^{2}u)^{-1}$ with $u=s^{2}$, $\int_{0}^{1}s^{2}ds=\frac13$) gives $\frac{\arctanh r}{r}>\frac{3}{3-r^{2}}$.
Therefore, we have $(3-r^{2})f(r)>3(1-r^{2})$. The coefficient of $Z^{2}$ in $B_{r}(Z)$ is
\begin{align}
    \alpha_{r}
    =
    \frac{(3-r^{2})f(r)-3(1-r^{2})}{2(1-r^{2})}>0.
\end{align}
Moreover, direct computation gives $B'_{r}(Z)=2\alpha_{r}\left(Z-\frac{1}{r}\right)$. Since $r<1$, one has $\frac{1}{r}>1$.
Thus $B'_{r}(Z)\leq0$ on $[-1,1]$, and consequently $B_{r}(Z) \geq B_{r}(1)$. It remains to prove $B_{r}(1)\,\geq\,0$.

Let $L(r)=\arctanh r$.
A direct computation gives $B_{r}(1)=\frac{1-r}{r^{4}}J(r)$, where $J(r)=(r^{3}+r^{2}-r-1)L(r)^{2}-(r^{3}+r^{2}-2r)L(r)+2r^{3}-r^{2}$. Furthermore, $J^{(4)}(r)=\frac{16\left(2(1+r)^{2}L(r)+r^{2}-2r\right)}{(1-r)^{3}(1+r)^{4}}$. 
Since $\arctanh r \geq r$, we have
\begin{align}
    2(1+r)^{2}L(r)+r^{2}-2r
    \geq2(1+r)^{2}r+r^{2}-2r
    =5r^{2}+2r^{3}
    \geq0.
\end{align}
Hence $J^{(4)}(r)\geq0$ for all $r\in[0,1)$.
Since $J(0)=J'(0)=J''(0)=J'''(0)=0$, we obtain $J(r)\geq0$ for $r<1$.
Therefore we have that $B_{r}(1)\geq0$.

\noindent{\bf Proof of (3):}

Combining the above two inequalities gives
\begin{align}
    \lambda_{-}(M^{-1}N) \geq \frac{p_{1}}{\bar{q}\,\widehat{q}}\,\bar{q}^{2}=\frac{\bar{q}\,p_{1}}{\widehat{q}}=\frac{p_{1}(1-p_{2})}{p_{2}(1-p_{1})}.
\end{align}
It remains to verify that equality can be attained. 

On the one hand, equality in~\eqref{eq:eb} holds if and only if $R=0$, i.e., $x^{2}+z^{2}=1$.

On the other hand, when $w_{1}=0$, both
$\mathcal{M}(\vec{w})$ and
$D_{q}\mathcal{M}(\vec{w}_{\mathcal{A}_{q}})D_{q}$
are diagonal. Hence the two eigenvalues of
\begin{align}
    \mathcal{M}(\vec{w})^{-1}
    D_{q}\mathcal{M}(\vec{w}_{\mathcal{A}_{q}})D_{q}
\end{align}
are $\mu_y$ and $\bar{q}^{2}$, corresponding
to the $w_1$- and $w_3$-directions,
respectively. Since $\mu_y\geq\bar{q}^{2}$
by Appendix~\ref{ap:D}, the smaller eigenvalue
is $\bar{q}^{2}$.

At $\vec{w}_{*}=(0,0,2p_{1}-1)$, which corresponds to $(x,z)=(0,-1)$ and hence $R=0$, we therefore obtain
\begin{align}
    \lambda_{-}(\vec{w}_{*})
    &=\frac{p_{1}}{(1-q)\widehat q}(1-q)^{2}=\frac{p_{1}(1-q)}{\widehat q}=\frac{p_{1}(1-p_{2})}{p_{2}(1-p_{1})}.
\end{align}

\end{document}